\documentclass[journal, twoside]{IEEEtran}

\ifCLASSINFOpdf
\usepackage[pdftex]{graphicx}
  \else
  \fi
\usepackage{cite}

\usepackage{graphicx}
\usepackage{footnote}
\usepackage{bbding}
\usepackage{pifont}

\usepackage[table,xcdraw]{xcolor}
\usepackage{diagbox}
\usepackage[flushleft]{threeparttable}
\makesavenoteenv{tabular}
\usepackage{tabu}
\usepackage{longtable}
\usepackage{epstopdf}
\usepackage{enumitem}
\usepackage{color}
\usepackage{framed}
\usepackage[ruled,linesnumbered]{algorithm2e}
\usepackage{algorithmic}
\usepackage{amssymb,amsmath}
\usepackage{amsmath}
\usepackage{mathrsfs}
\usepackage{amsmath,amsthm,amssymb,amsfonts}
\usepackage{multirow}
\usepackage{booktabs}
\usepackage{subfigure}
\usepackage{array}
\usepackage{makecell}
\usepackage{stfloats}
\usepackage{epstopdf}
\usepackage{epsfig}
\usepackage{bm}
\usepackage{cite}

\setlist{parsep=0pt,listparindent=\parindent}

\graphicspath{ {image/} }

\ifCLASSINFOpdf

\else

\fi

\begin{document}

\title{Unsupervised Euclidean Distance Attack on Network Embedding}

\author {Shanqing~Yu,
        Jun~Zheng,
        Jinhuan~Wang,
        Jian~Zhang,\\
        Lihong~Chen,
        Qi~Xuan,~\IEEEmembership{Member,~IEEE,}
        Jinyin~Chen,
        Dan~Zhang,\\
        and Qingpeng~Zhang,~\IEEEmembership{Member,~IEEE}

\thanks{This work is partially supported by National Natural Science Foundation of China (61572439, 61502423), Natural Science Foundation of Zhejiang Province(LR19F030001, LY19F020025), and Key technologies, system and application of Cyberspace Big Search, Major project of Zhejiang Lab (2019DH0ZX01).
(\emph{Corresponding author: Shanqing Yu.})}

\thanks{S. Yu, J. Zheng, J. Wang, J. Zhang, L. Chen, Q. Xuan, J. Chen, and D. Zhang are with the
Institute of Cyberspace Security, and the college of information engineering, Zhejiang University of Technology, Hangzhou 310023, China (e-mail: yushanqing@zjut.edu.cn; 490728335@qq.com; JinhuanWang@zjut.edu.cn; zj\_1994@outlook.com; xuanqi@zjut.edu.cn; coffeeclh@163.com; chenjinyin@zjut.edu.cn; danzhang@zjut.edu.cn).}

\thanks{S. Yu, Q. Xuan, and J. Chen are also with the Zhejiang Lab, Hangzhou 311121, China.}

\thanks{Q. Zhang is with the City University of Hong Kong, Hong Kong, China (e-mail:  qingpeng.zhang@cityu.edu.hk).}}

\markboth{}%
{}

\maketitle

\begin{abstract}
Considering the wide application of network embedding methods in graph data mining, inspired by the adversarial attack in deep learning, this paper proposes a Genetic Algorithm (GA) based Euclidean Distance Attack strategy (EDA) to attack the network embedding, so as to prevent certain structural information from being discovered.
EDA focuses on disturbing the Euclidean distance between a pair of nodes in the embedding space as much as possible through minimal modifications of the network structure.
Since a large number of downstream network algorithms, such as community detection and node classification, rely on the Euclidean distance between nodes to evaluate the similarity between them in the embedding space, EDA can be considered as a universal attack on a variety of network algorithms.
Different from traditional supervised attack strategies, EDA does not need labeling information, and, in other words, is an unsupervised network embedding attack method.

We take DeepWalk as the base embedding method to develop the EDA. Experiments with a set of real networks demonstrate that the proposed EDA method can significantly reduce the performance of DeepWalk-based networking algorithms, i.e., community detection and node classification, outperforming other attack strategies in most cases.
We also show that EDA works well on attacking the network algorithms based on other common network embedding methods such as High-Order Proximity preserved Embedding (HOPE) and non-embedding-based network algorithms such as Label Propagation Algorithm (LPA) and Eigenvectors of Matrices (EM). The results indicate a strong transferability of the EDA method.
\end{abstract}


\maketitle

\IEEEpeerreviewmaketitle

\ifCLASSOPTIONcompsoc

\IEEEraisesectionheading{\section{Introduction}\label{sec:introduction}}
\else
\section{Introduction}
\label{sec:introduction}
\fi
\IEEEPARstart{R} eal-world complex systems can be represented and analyzed as networks. During the past few decades, network science has emerged as an essential interdisciplinary field aiming at using network and graph as a tool to characterize the structure and dynamics of complex systems, including social networks, citation networks, protein networks and transport networks~\cite{barabasi2016network}.
Quite recently, numerous methods have
been proposed to map the nodes of a network into vectors in an Euclidean space, namely network embedding, which largely facilitates, the application of machine learning methods in graph data mining~\cite{bengio2013representation,hong2019deep}.
Network embedding solves the problem of high dimensional and sparseness of the original network data. It builds a bridge between machine learning and network science, enabling many machine learning algorithms to be applied in network analysis.

However, while providing convenience to users, such network analysis algorithms may also bring the risk of privacy leakage, i.e., our personal information in the social network may be easily predicted by adopting such algorithms~\cite{liben2007link,andersen2006local,fortunato2010community}. In recent studies of deep learning, it is found that, while performing exceptionally well in a variety of tasks, deep neural networks seem to be susceptible to small perturbations, especially in the area of computer vision~\cite{szegedy2013intriguing,athalye2017synthesizing,papernot2016limitations,kurakin2016adversarial}. These adversarial attacks usually target at specific algorithms and make the prediction accuracy drop sharply. Quite recently, it was also found that network algorithms in community detection~\cite{chen2018ga}, link prediction~\cite{yu2018target}, and node classification~\cite{zugner2018adversarial,chen2018fast,wang2018attack} are also vulnerable to such adversarial attacks.

Recently, researchers have invested more and more energy to the study of adversarial attacks in network algorithms. The main applications and motivations for these works have been considered as follows. Firstly, although the network algorithm has developed rapidly and has been recognized by many people, the robustness of its has not been widely verified, which part is indispensable before generally applied. The second is about the protection of private information that is predicted in social networks. Thirdly, it is exactly because of the existence of adversarial attacks research, which is more conducive to the development and advancement of network algorithms.

Network algorithms, especially machine learning-based network algorithms,  also have potential security vulnerabilities and are vulnerable to attacks. For example, the purpose of the adversarial attacks for the node classification is to make the target node be misclassified, and hides the attribute of nodes; for the community detection algorithms attacks, the sample of adversarial networks can break community prediction down, and disorder of the community function. The attacks against the network embedding method will make the model generate the wrong node vectors, which result in the blindness of downstream algorithms based on the representation calculation.

All of those research in network algorithms attack can be transferred from the adversarial study in the field of computer vision by perturbing a tiny number of nodes or links into the target network. In terms of applications, the adversarial attack strategy of network algorithms seems to be a disturbance for network analysts; On the other hand, for Internet users, it can be considered as a privacy protection strategy, which can reduce the risk of information leakage. We also believe that competition in adversarial learning can provide promotion and improvement to network algorithms. The attack and defense will also become an essential part in further research.

In this paper, we focus on attacking the network embedding process, rather than some particular network algorithms.
Since most network algorithms are based on network embedding, attacking the embedding process (instead of particular algorithms) could be a more generic approach that can be easily applied to various attack tasks.
In this consideration, here we propose an Euclidean Distance Attack (EDA), aiming to disturb the distances between vectors in the embedding space directly. We think it is the vector distance that determines the performance of many downstream algorithms based on network embedding. Since we do not need to know the labels of training data, it can be considered as an unsupervised learning method. In particular, the main contributions of this paper are as below:
\begin{itemize}
\item We propose a new unsupervised attack strategy on network embedding, namely EDA, using the Euclidean Distance between embedding vectors as the reference.
\item We adopt the Genetic Algorithm (GA) to search for the optimal solution for EDA. As compared with state-of-the-art baseline attack strategies, EDA performed the best in reducing the prediction accuracy of downstream algorithms for community detection and node classification tasks.
\item We validate the transferability of EDA in reducing the performance on many other network algorithms.
\end{itemize}

\hspace*{\fill} \\
The rest of the paper is organized as follows. In Sec.~\ref{sec:relatedwork}, we introduce the related work of network embedding and adversarial attacks. In Sec.~\ref{sec:methodology}, we introduce our proposed attack method. In Sec.~\ref{sec:experiment}, we compare the performance of EDA and other baseline strategies on attacking community detection and node classification algorithms in four real-world networks. Finally, we conclude the paper with discussions of future work in Sec.~\ref{sec:conclusion}.

\section{Related work}
\label{sec:relatedwork}

\subsection{Network embedding}
With the development of machine learning technology, the feature learning for nodes in the network has become an emerging research task.
The embedding algorithm transforms network information into the low-dimensional dense vector, which is input for machine learning algorithms.

In the recent study, Perozzi et al.~\cite{perozzi2014deepwalk} proposed the first network embedding method, namely DeepWalk, which introduced Natural Language Processing (NLP) model~\cite{collobert2008unified,mikolov2013efficient,mikolov2013distributed} into network and achieved great success in community detection~\cite{fortunato2016community,guerrero2017adaptive,tang2011leveraging,liu2018we} and node classification~\cite{tang2016node,bhagat2011node}.
Grover et al.~\cite{grover2016node2vec} developed node2vec as an extension of DeepWalk. They utilized a biased random walk to combine BFS and DFS neighborhood exploration so as to reflect equivalence and homogeneity in the network structure.
Moreover, Tang et al.~\cite{tang2015line} presented LINE that preserved both the first-order and second-order proximities to rich node representation.
For some specific tasks, such as community detection, Wang et al.~\cite{wang2017community} proposed a new Modularized Nonnegative Matrix Factorization (M-NMF) model that incorporates community structures into network embedding. Their approach is to jointly optimize the NMF-based representation learning model and the module-based community detection model so that the learning representation of the node can maintain both microscopic and community structures.
Lai et al.~\cite{NIPS2017_7110} formulated a multi-task Siamese neural network structure to connect embedding vectors and combine the global node ranking with local proximity of nodes. This model can more abundantly retain the information from the network structure.

These network embedding methods show the following advantages. First, embedding can well obtain the local structural information of the network; second, many methods can be easily parallel so as to decrease the computation complexity. The representation vectors of nodes, obtained by the network embedding algorithms, can be used to support the subsequent network analysis tasks, such as link prediction, community detection, and node classification.

\subsection{Adversarial attacks}
Given the importance of graph analysis, increasing number of works start to analyze the robustness of machine learning models on graph.
For instance, due to the connectivity and cascading effects in networks, Faramondi et al.~\cite{faramondi2018network} analyze the robustness and vulnerability of complex networks.
Waniek et al.~\cite{waniek2018hiding} developed a heuristic method, namely Disconnect Internally, Connect Externally (DICE). They added perturbations to the original network by deleting the links within community while adding the links between communities. Chen et al.~\cite{chen2018ga} proposed an attack strategy against community detection, namely $Q$-Attack, which uses modularity $Q$, under certain community detection algorithm, as the optimization objective, aiming to disturb the overall community structure. Yu et al.~\cite{yu2018target} introduced an evolutionary method against link prediction using Resource Allocation Index (RA) so as to protect link privacy. As for node classification, Z{\"u}gner et al.~\cite{zugner2018adversarial} proposed NETTACK to fool Graph Convolutional Networks (GCN) by generating adversarial networks. Chen et al.~\cite{chen2018fast} further proposed Fast Gradient Attack (FGA), which utilized the gradient of algorithm to design loss function. FGA model can generate adversarial network faster and make the target nodes classified incorrectly with quite small perturbations. Moreover, Wang et al.~\cite{wang2018attack} designed a greedy algorithm to attack GCNs by adding fake nodes. This method generated adjacency and feature matrices of fake nodes to minimize the classification accuracy of the existing nodes.
Those kinds of attack strategies are always effective in most cases since they are strongly targeted at specific algorithms.

Most of the current attack strategies belong to supervised learning, i.e., attackers can get actual labels of nodes or communities, and further utilize this information to design attack strategies. However, in many cases, it is difficult and costly to collect such information, making those supervised attack strategies less effective.
In order to solve these problems, Bojcheski et al.~\cite{bojcheski2018adversarial} analysis adversarial vulnerability on the widely used models based on random walks firstly and derive efficient adversarial perturbations that poison the network structure.
What's more, Sun et al.~\cite{sun2018data} propose an efficient method based on projected gradient descent to attack unsupervised node embedding algorithms.

\section{methodology}
\label{sec:methodology}
In this section, we introduce \emph{DeepWalk} briefly, based on which we propose the Euclidean distance attack (EDA) method. In particular, we turn the attack problem to a combinatorial optimization problem, and then use Genetic Algorithm (GA) to solve it. Here, we choose DeepWalk because it is one of the most widely-used unsupervised network embedding methods and it can have good mathematical properties rooted at matrix factorizations~\cite{qiu2018network}.

\subsection{DeepWalk}
This paper mainly focuses on undirected and unweighted networks. A network is represented by $G(V, E)$, where $V$ denotes the set of nodes and, $E$ denotes the set of links. The link between nodes $v_i$ and $v_j$ is denoted by $e_{ij} = (v_i,v_j) \in E$. The adjacency matrix of the network then is defined as $A$, with the element denoted by
    \begin{equation}
        a_{ij}=\left\{
        \begin{array}{ll}
        1 \quad (v_i,v_j) \in E\\
        0 \quad (v_i,v_j) \notin E.
        \end{array}\right.\label{Eq:matixA}
    \end{equation}
Real-world networks are often sparse and high-dimensional, preventing the broad application of machine learning algorithms in graph data. To address these problems, network embedding is a family of methods to encode the topological properties in the graph data into low-dimensional features.

DeepWalk trains the vectors of nodes $R^{|V|\times n}$ by calculating the probability of generating the nodes on both sides from the center node, with the loss function represented by
    \begin{eqnarray}
         \mathop{min}\limits_{R} \sum\limits_{k = -w,k\neq 0}^w{-\log P(v_{i+k} \mid v_i)}, \label{Eq:deepwalkloss}
    \end{eqnarray}
where the sequence $\left\{v_{i-w},\cdots,v_{i-1},v_{i+1},\cdots,v_{i+w}\right\}$ is obtained by random walk within the window $w$ around the center node $v_i$, and the probability $P(v_{i+k}|v_i)$ can be transformed by the following softmax function~\cite{mnih2009scalable,morin2005hierarchical}:
    \begin{eqnarray}
         P(v_{i+k} \mid v_i) = \frac{\exp(r_ir^{T}_{i+k})}{\sum\limits_{n = 1}^{|V|}{\exp(r_ir^{T}_{n})}},\label{Eq:deepwalkpr}
    \end{eqnarray}
where $r_i$ is the representation vector of node $v_i$.

\subsection{EDA on network embedding}
Many machine learning methods are based on the relative, rather than absolute, positions of samples in Euclidean space.
Thus the Euclidean distance between samples playing a vital role in these methods. Moreover, due to the randomness of many network embedding algorithms, embedding vectors generated for the same node in different rounds might be different from each other.
Regardless of such differences, the Euclidean distance between the same pair of nodes in the embedding space is approximately consistent.
Those are the key motivation that drives us to propose the Euclidean Distance Attack (EDA): \emph{Disturbing the Euclidean distance between pairwise nodes in the embedding space as much as possible with the minimal changes of the network structure.}

    \begin{figure*}[!t]
    	  \centering
     	  \includegraphics[width=15cm,height=4.2cm]{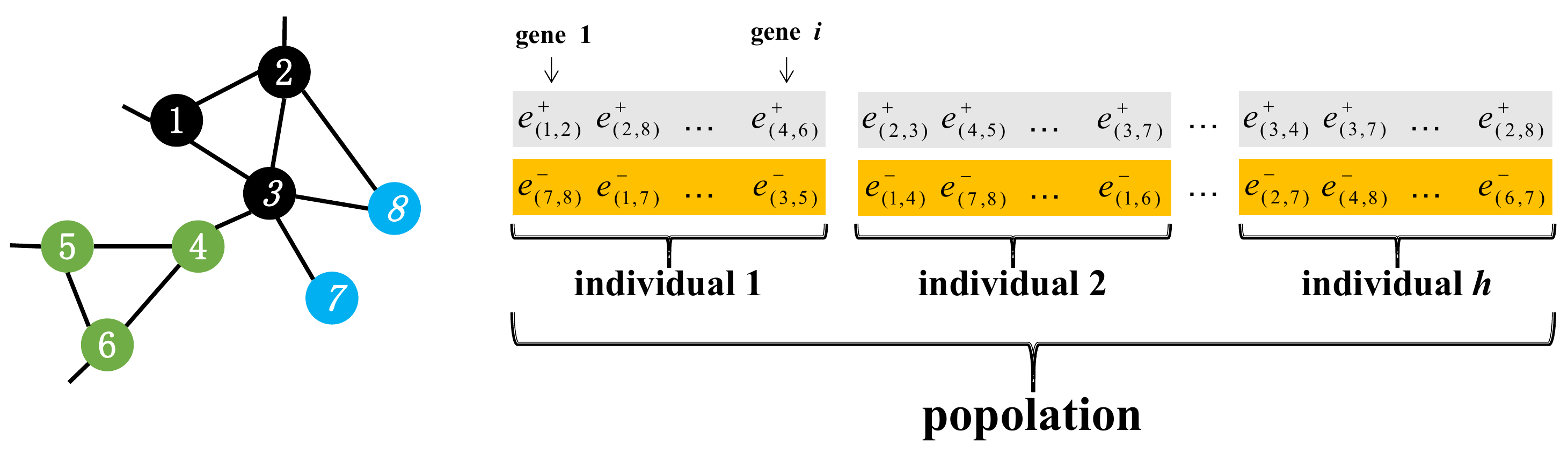}
    	  \caption{Encoding of network. The individual solution includes $i$ genes formed by adding and deleting links, and a population consists of $h$ individuals. The sets of added and deleted links are represented in gray and yellow, respectively.}
          \label{Eq:encoding}
    \end{figure*}

In particular, we calculate the distance between a pair of nodes $v_i$ and $v_j$ in the embedding space as follows:
\begin{eqnarray}
    d_{ij}= \mathbf{dist}(r_{i},r_{j}) = ||r_{i}-r_{j}||_2,
\label{Eq:dij}
\end{eqnarray}
based on which we can get the Euclidean distance matrix $D=[d_{ij}]_{|V|\times{|V|}}$ for the whole network, with each row denoted by $D_i$ representing the Euclidean distances between node $v_i$ and all the other nodes in the network.

Denote the adversarial network after our EDA as $\hat{G}$, and its corresponding Euclidean distance matrix in the embedding space as $\hat{D}$. We calculate the Pearson correlation coefficient between the distance vectors of the corresponding nodes in the original network and adversarial network:
\begin{eqnarray}
    \varphi(G,\hat{G})= \sum\limits_{i=1 }^{|V|}\bm|{\rho}(D_{i},\hat{D}_{i})|.\label{Eq:Dg}
    \end{eqnarray}
Then, we focus on minimizing $\varphi$ by changing a certain number of links in the network with the following objective function:
    \begin{eqnarray}
        \min\varphi(G,\hat{G}).
    \end{eqnarray}

    \begin{algorithm}[!b]
        \caption{Method of EDA}
        \label{alg:litm}
        \KwIn{\textit{The original network} $G=(V,E)$}
        \KwOut{ \textit{Adversarial network $\hat{G}^\ast$}}
        Initialize \textit{the node vectors} $R^{|V|\times n}$ \textit{and the distance matrix} $D$, with $d_{ij} = \mathbf{dist}(r_{i},r_{j})$;

        \While{not converged}
        {
          $\hat{R}^{|V|\times n} = \mathbf{DeepWalk}(\hat{A},n)$;

          \For{$i=1;i \le |V|;i++$}
          {
          \For{$j=1;j < i;j++$}
          {
            $\hat{d_{ij}} = \sqrt{\sum\limits_{q = 1}^{n}(\hat{r^{q}_{i}}-\hat{r^{q}_{j}})^{2}}$;
          }
          }

          \For{$i=1;i \le |V|;i++$}
          {

            ${\rho}_{\tau} += \mathbf{\rho}(D_{i},\hat{D_{i}})$\;

          }
          $fitness = 1 - \frac{{\rho}_{\tau}}{|V|}$ \;
           $\hat{G} = \mathbf{Genetic\,Algorithm}(G,fitness)$
        }
        return \textit{Adversarial network} $\hat{G}^\ast$;
    \end{algorithm}

\subsection{Process of perturbation}

Usually, we want the perturbation to be small enough to make the attack imperceptible for others.
Taking this into consideration, we add or remove very low proportions of the links on the original network.
What's more, there are many forms of perturbation in networks. We can also implement the attack as a rewiring process, i.e., adding a new link while deleting an existent one at the same time, so that the total number of links remains the same after the attack.

In process of perturbation, we denote the set of added links as $E^+\subseteq{\overline{E}}$ and the set of deleted links as $E^-\subseteq{E}$, where $\overline{E}$ is the set of all pairs of unconnected nodes in $G$. Then, we get the adversarial network $\hat{G}(V,\hat{E})$ with the updated set of links $\hat{E}$ satisfying
\begin{eqnarray}
    \hat{E} = E\cup E^+ - E^- \label{Eq:Ek}
\end{eqnarray}

Suppose $u$ is the number of flipped links in the attack, the total complexity of instances in the searching space is equal to
\begin{eqnarray}
    O(u)= \mathcal{C}_{|E|}^{u} *  \mathcal{C}_{|\overline{E}|}^{u}, \label{Eq:Tk}
\end{eqnarray}
which could become huge as the size of the network or the number of flipping links grows. The search for the optimal solution is NP, and thus we adopt the Genetic Algorithm (GA) to search for the optimal solution. The detailed procedure of EDA is presented in Algorithm~\ref{alg:litm}.

\subsection{The evolution of EDA}

Here, we use the GA to find the optimal set of flipping links for EDA. Typically, the algorithm consists of three parts: encoding of the network, selection by fitness, crossover and mutation operation:
\begin{itemize}
    \item \bfseries Encoding of network: \mdseries We directly choose the flipped links as genes, including the set of deleted links $E^-$ and the set of added links $E^+$. The length of each chromosome is equal to the number of added or deleted links. Fig.~\ref{Eq:encoding} shows an overview of network encoding. Individuals are combinations of flipping links, representing different solutions of adversarial perturbations, and a population consists of $h$ individuals.
    \item\bfseries Selection by fitness: \mdseries We use Eq.~(\ref{Eq:fitness}) as the fitness function of individual $k$ in GA, capturing the relative changes of vector distances in the embedding space by the attack:
    \begin{eqnarray}
        f(k)  = 1- \frac{\varphi(G,\hat{G})}{|V|}.
     \label{Eq:fitness}
    \end{eqnarray}
    Then, the probability that individual $k$ is selected to be the parent genes in the the next generation is proportional to its fitness:
    \begin{equation}
        p(i) = \frac{f(i)}{\sum_{j=1}^{h}f(j)}.
    \end{equation}

\item \bfseries Crossover and mutation: \mdseries We then use the selected individuals of higher fitness as the parents to generate new individuals by adopting crossover and mutation operations, assuming that those better genes can be inherited in the process. In particular, single point crossover between two individuals is used, with probability $p_c$, as shown in Fig.~\ref{Eq:Crossover}; while for mutation, we randomly select individuals from a population and randomly change their genes, with probability $p_m$, as shown in Fig.~\ref{Eq:mutation}. 


    \begin{figure}[!t]
    	  \centering
     	  \includegraphics[width=\linewidth]{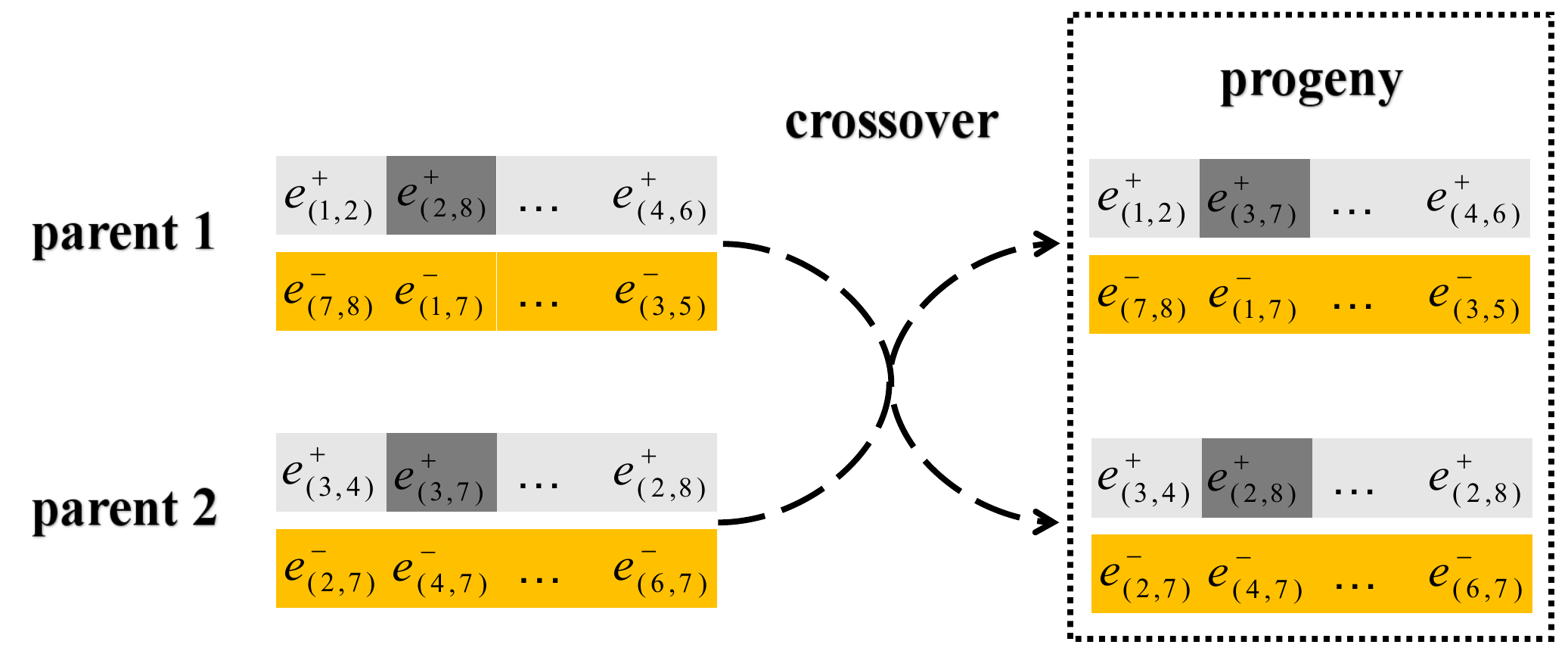}
    	  \caption{An example of crossover operation, where link $(2,8)$ in parent 1 and link (3,7) in parent 2 are exchanged to produce progeny.}
          \label{Eq:Crossover}
    \end{figure}

    \begin{figure}[!t]
    	  \centering
     	  \includegraphics[width=\linewidth]{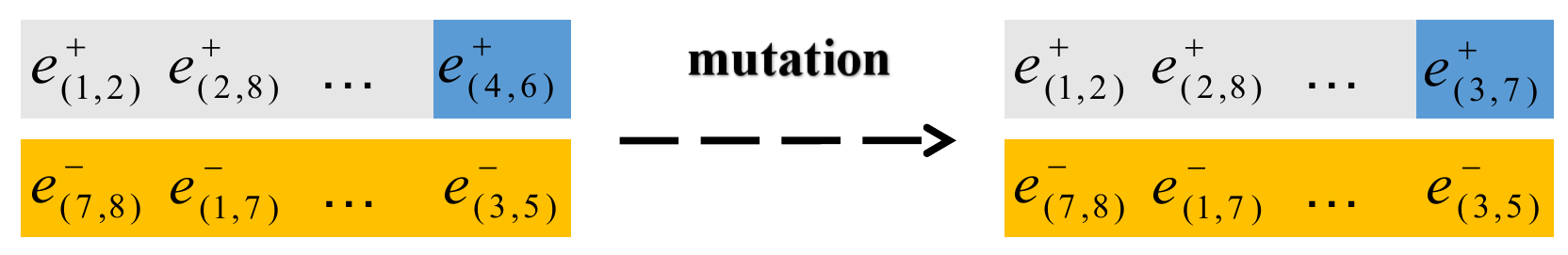}
    	  \caption{An example of crossover operation£¬ where a new link (3,7) is generated  by mutation.}
          \label{Eq:mutation}
    \end{figure}

  \begin{algorithm}[!t]
        \caption{Genetic Algorithm (GA)}
        \label{alg:GA}
        \KwIn{\textit{The fitness of individuals} $f(k), \ k=1,2,\dots,h$ \textit{and the parameters} $h,\  p_c,\  p_m$;}
        \KwOut{\textit{Adversarial network $\hat{G}$}}
        Initialize \textit{individuals in population};

          $Elites = \mathbf{Retain}(f(k),population) $;\
          $Selection = \mathbf{Selection}(f(k),individual)$;\
          $Crossover = \mathbf{Crossover}(p_c,Section)$;\
          $Mutation = \mathbf{Mutation}(p_m,Section)$\;

          population = $Elites \cup Crossover \cup Mutation$;\

        \textit{Reconstruct network}.\\
        return \textit{Evolutionary network} $\hat{G}$;
    \end{algorithm}

\end{itemize}

\subsection{Overview of EDA}
\begin{figure*}[t]
    	  \centering
     	  \includegraphics[width=\linewidth]{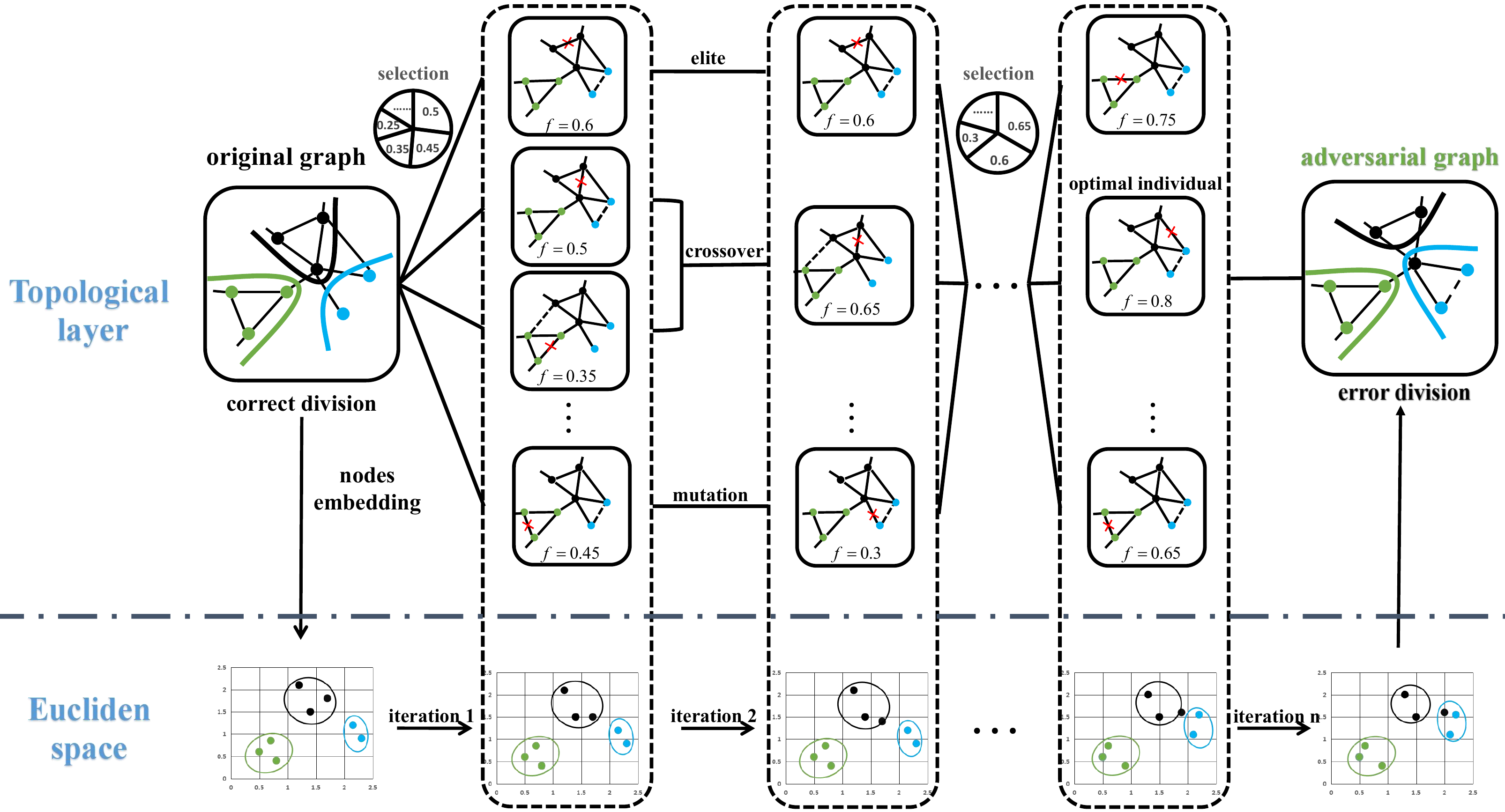}
    	  \caption{The framework of EDA on a network. With the evolution of perturbation, the positions of nodes in the Euclidean space constantly changes, leading to the changing prediction results of node classification.}
          \label{Eq:method1}
\end{figure*}

In the Fig.~\ref{Eq:method1}, we divided the whole generating of adversarial network in EDA into two parts, the topological layer and the Euclidean space layer.
In the topological layer, we perturb the original network by adding links or deleting links, which can be observed directly through the figure.
In the lower part of Fig.~\ref{Eq:method1}, it is the Euclidean space layer, which is obtained by mapping the network structure through the graph embedding algorithm. We calculate the change in Euclidean space and choose the fliped links.
EDA method uses the Genetic Algorithm to iterate the initially generated perturbution, and finally blind the prediction from graph algorithms.

As shown in Fig.~\ref{Eq:method1}, the black node in the original network begins to be divided into the upper part of the community by the community detection algorithm based on the embedding model. However, after $n$ times of evolutionary iterations, by flipping some links, the node is incorrectly divided into the lower part of the community. Our strategy utilizes Genetic Algorithm to generate the optimal adversarial network through selection, crossover and mutation.

\section{EXPERIMENTS}
\label{sec:experiment}
In order to evaluate the effectiveness of EDA, we compare it with four baseline methods by performing multi-task experiments on several real-world networks.

\subsection{Data sets}

\begin{table}[h]\renewcommand{\arraystretch}{1.2}
\setlength{\tabcolsep}{6mm}
	\newcommand{\tabincell}[2]{\begin{tabular}{@{}#1@{}}#2\end{tabular}}
	\caption{Basic structural features of three networks.
             $|V|$ and $|E|$ are the numbers of nodes and links, respectively;
             $\langle k\rangle$ is the average degree; $C$ is the clustering coefficient
             and $\langle d \rangle$ is the average distance.}
    \centering
	\begin{centering}

		\begin{tabular}{c|cccc}

			\Xhline{1pt}

			& Karate & Game & Citeseer & Cora\tabularnewline
			\Xhline{1pt}
			$|V|$& \tabincell{c}{34} & 107 & 3312 & 2708\tabularnewline
			$|E|$& \tabincell{c}{78} & 352 & 4732 & 5429\tabularnewline
			$\langle k\rangle$& \tabincell{c}{6.686} & 6.597 & ** & **\tabularnewline
			$C$& \tabincell{c}{0.448} &0.259 & ** & **\tabularnewline
			$\langle d \rangle$ & \tabincell{c}{2.106} & 2.904 & ** & **\tabularnewline
			\Xhline{1pt} 			
		\end{tabular}		
	\end{centering}	
	\label{topo}
\end{table}

Here, we choose four commonly-used benchmark social networks to evaluate the performance of the attack strategies.
\begin{itemize}
\item \textbf{Zachary Karate Club (Karate)}~\cite{ghosh2010community}. The karate network is constructed by observing a university karate club, which consists of $34$ nodes and $78$ links. 

\item \textbf{Game of Thrones (Game)}~\cite{beveridge2016network}. The TV sensation \emph{Game of Thrones} is based on George Martin's epic fantasy novel series \emph{A Song of Ice and Fire}. It is a network of character interactions in the novel, which contains $353$ links connecting $107$ characters.

\item \textbf{The citeseer dataset (Citeseer)}. This dataset is a paper citation network with nodes are divided into six classes ~\cite{mccallum2000automating}. $3,312$ papers and $4,732$ citation links are in this network.

\item \textbf{The cora dataset (Cora)}. This dataset contains seven classes by combining several machine learning papers which contains $2,708$ papers and $5,429$ links in total ~\cite{mccallum2000automating}. The links between papers represent the citation relationships.

\end{itemize}

The basic topological properties of these networks are presented in TABLE~\ref{topo}.

\subsection{Baseline methods}\label{sec:baselines}
We perform experiments on community detection and node classification to see how the proposed EDA degrades their performances. In particular, we compare the performance of EDA with that of the following baseline methods.
		\begin{itemize}

          \item \textbf{Randomly Attack (RA).} RA randomly deletes the existent links, while randomly adds the same number of nonexistent links. This attack strategy does not require any prior information about the network.

          \item \textbf{Disconnect Internally, Connect Externally (DICE).} DICE is a heuristic attack algorithm for community detection~\cite{waniek2018hiding}. The attacker needs to know the true node labels in the network and then delete the links within community and add the links between communities.

          \item \textbf{Degree-Based Attack (DBA).} It has been found that many real-world networks follow the power-law distribution~\cite{barabasi1999emergence}, in which a small fraction of nodes (usually named as hubs) have most connections. Since it is generally recognized that these hub nodes often have a huge impact on the connectivity of the network, here we also adopt another heuristic attack strategy named degree-based attack (DBA)~\cite{chen2018ga}. In each iteration, we select the node of the highest degree and delete one of its links or add a link to the node with the smallest degree. Then, we update the degrees of these nodes. DBA is only based on the structure of the network, but not on the labels and attributes of the nodes.

          \item \textbf{Greedy Attack (GDA).} GDA is a greedy algorithm based on our proposed EDA. Calculating the fitness of each link from the candidate set, and after sorting, selecting the Top-K links as perturbation. Even for greedy algorithms, the total number of flipped links is $|V|*(|V|-1)/2$, where $|V|$ is the number of nodes. If the number of nodes is higher than a thousand, the complexity is still millions. In our experiment, we take sampling to approximate the result. The $m$ links are randomly extracted from candidate set (where $m<<|V|*(|V|-1)/2$). Calculate the fitness value of each link, and select Top-K as the combined greedy solution.

          \item \textbf{Node Embedding Attack (NEA).} NEA is an attack strategy for network embedding proposed by Bojcheski et al.~\cite{bojcheski2018adversarial}. This method generates adversarial networks by maximizing the loss function of the adjacency matrix $\hat{A}$ and the matrix $Z^*$ embedded from the adversarial network.
        \end{itemize}

\subsection{Parameter setting and convergence of GA}
For DeepWalk and GA, there are many parameters. In our experiments, the parameter setting is empirically determined through balancing the performance and convergence speed shown in Table.~\ref{tab:parameter}. Note that different parameter settings may lead to various performances of these network algorithms, but our attack strategies will be still effective in degrading them.

\begin{figure}[!t]
\centering
     \subfigure[Karate network]{
        \includegraphics[width=.9\linewidth]{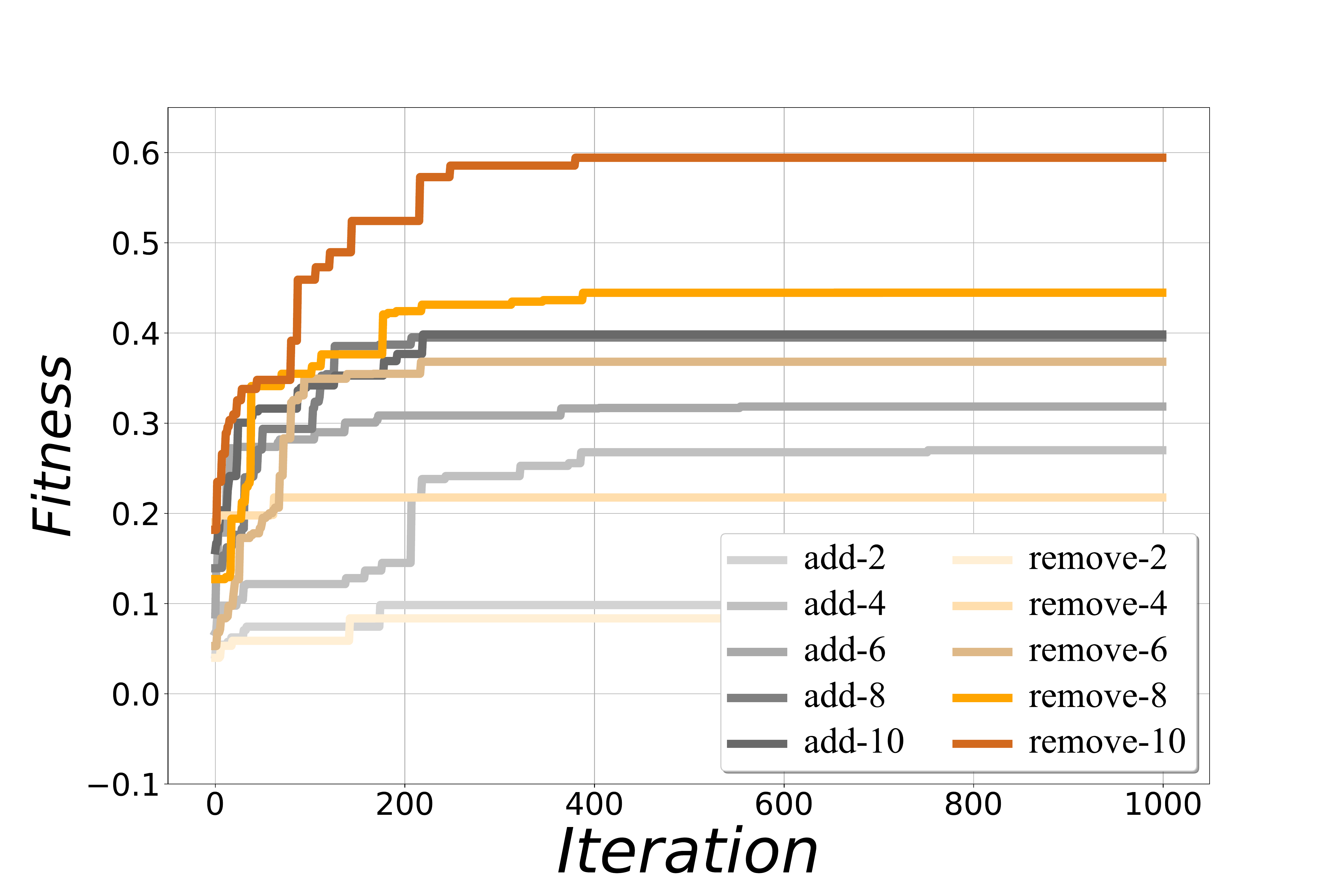}
    }
        \subfigure[Game network]{
        \includegraphics[width=.9\linewidth]{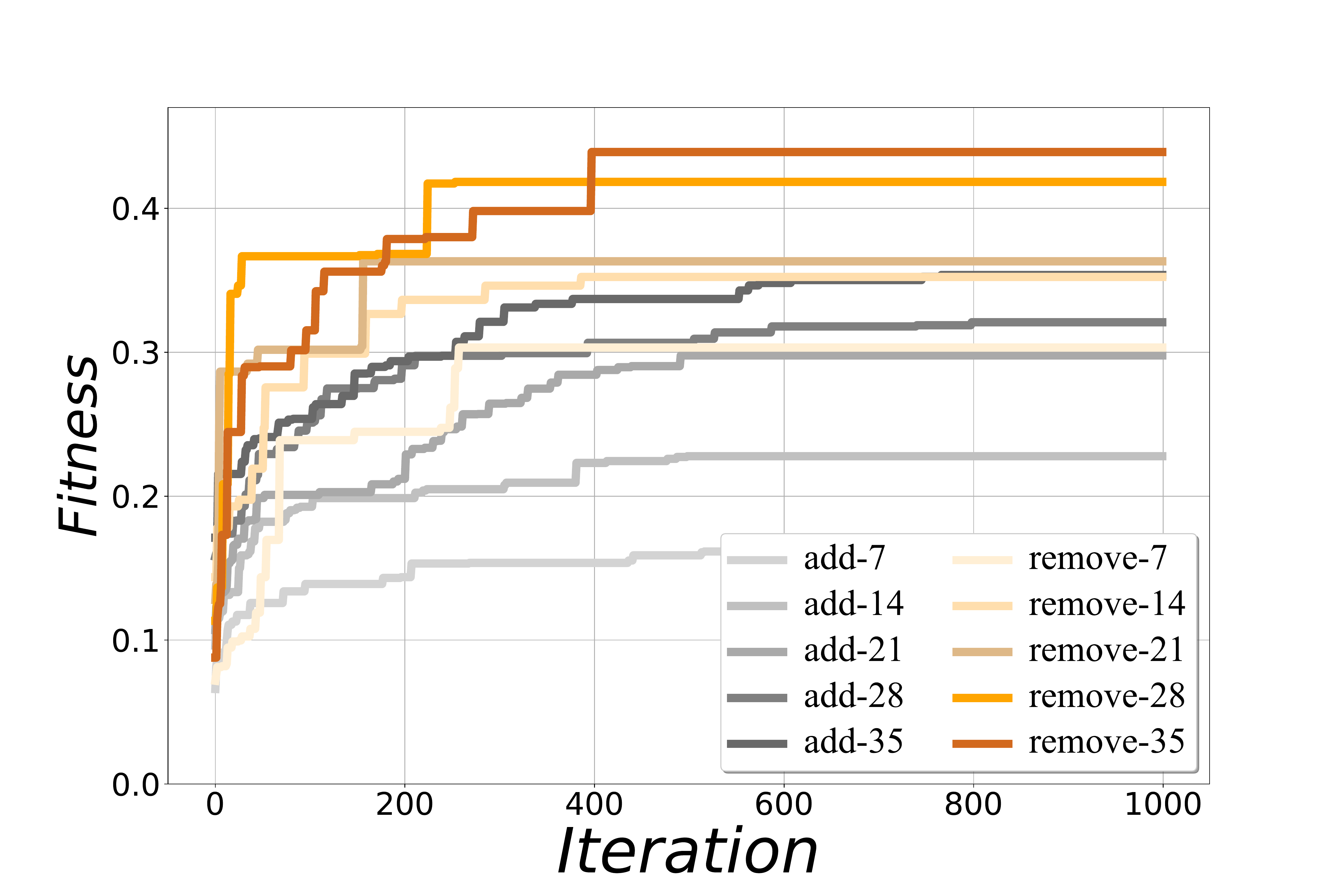}
    }
\caption{The above two figures are the EDA iteration diagrams of the karate network and the game network. The gray color represents the addition of the links, and the orange color represents the deletion of the links. }
 \label{Fig:convergence}
\end{figure}

Fig.~\ref{Fig:convergence} verifies the convergence of EDA. It can be seen from the experimental results of the genetic algorithm that, iteration can be converged after $500$ generations in most instances. What's more, deleting links can bring greater destruction to the network, compared to adding the same number of links.

\begin{table}[!t]\renewcommand{\arraystretch}{1}
\newcommand{\tabincell}[2]{\begin{tabular}{@{}#1@{}}#2\end{tabular}}
    \caption{Parameters setting for DeepWalk and GA.}
    \setlength{\tabcolsep}{3mm}
    \renewcommand\arraystretch{1.2}
    \centering
    \begin{centering}
    	\begin{tabular}{c|c|c}
    		\Xhline{1.2pt}
            $Item$& Meaning&Value \tabularnewline
            \Xhline{1.2pt}
                $n$& number of random walk&10 \tabularnewline
                $r$& size of window&5 \tabularnewline
                $w$& length of random walk&40 \tabularnewline
                $d$& size of representation&4/8/16/24 \tabularnewline
                \Xhline{0.6pt}
                $h$& number of population size&20 \tabularnewline
    			$n_{iteration}$& number of iterations& 1000 \tabularnewline
    			$n_{elite}$&  number of retained elites& 4\tabularnewline
    			$n_{crossover}$& number of chromosomes for crossover & 16  \tabularnewline
    			$n_{mutation}$ & number of chromosomes for mutation & 16 \tabularnewline
                $p_c$ & crossover rate&0.6 \tabularnewline
                $p_m$ & mutation rate&0.08 \tabularnewline
    		\Xhline{1.2pt}			
    	\end{tabular}		
    \end{centering}	
    \label{parameter} \label{tab:parameter}
\end{table}

\begin{figure*}[!t]
\centering
    \subfigure[Karate]{
        \includegraphics[width=.23\linewidth]{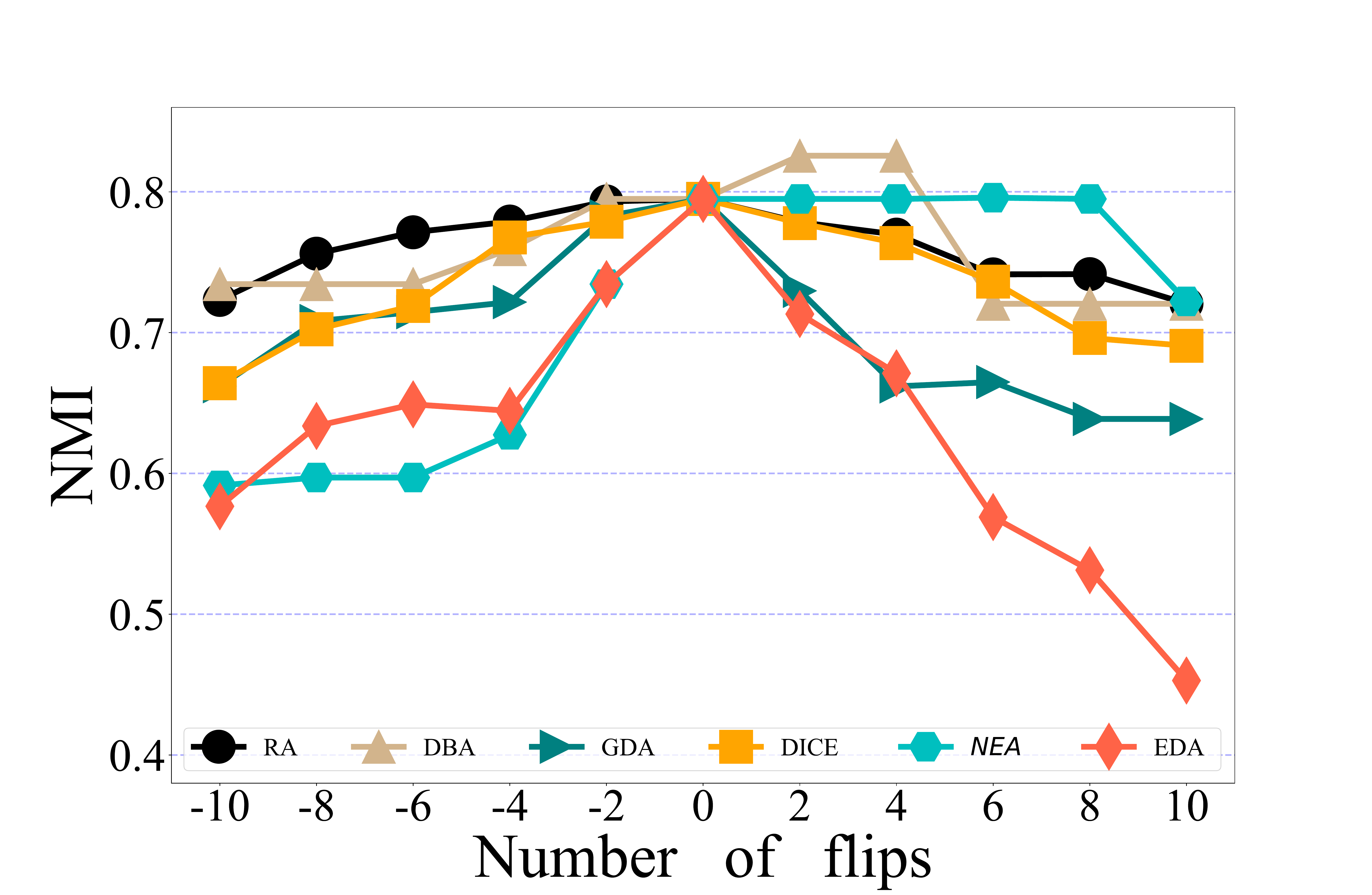}
    }
    \subfigure[Game]{
        \includegraphics[width=.23\linewidth]{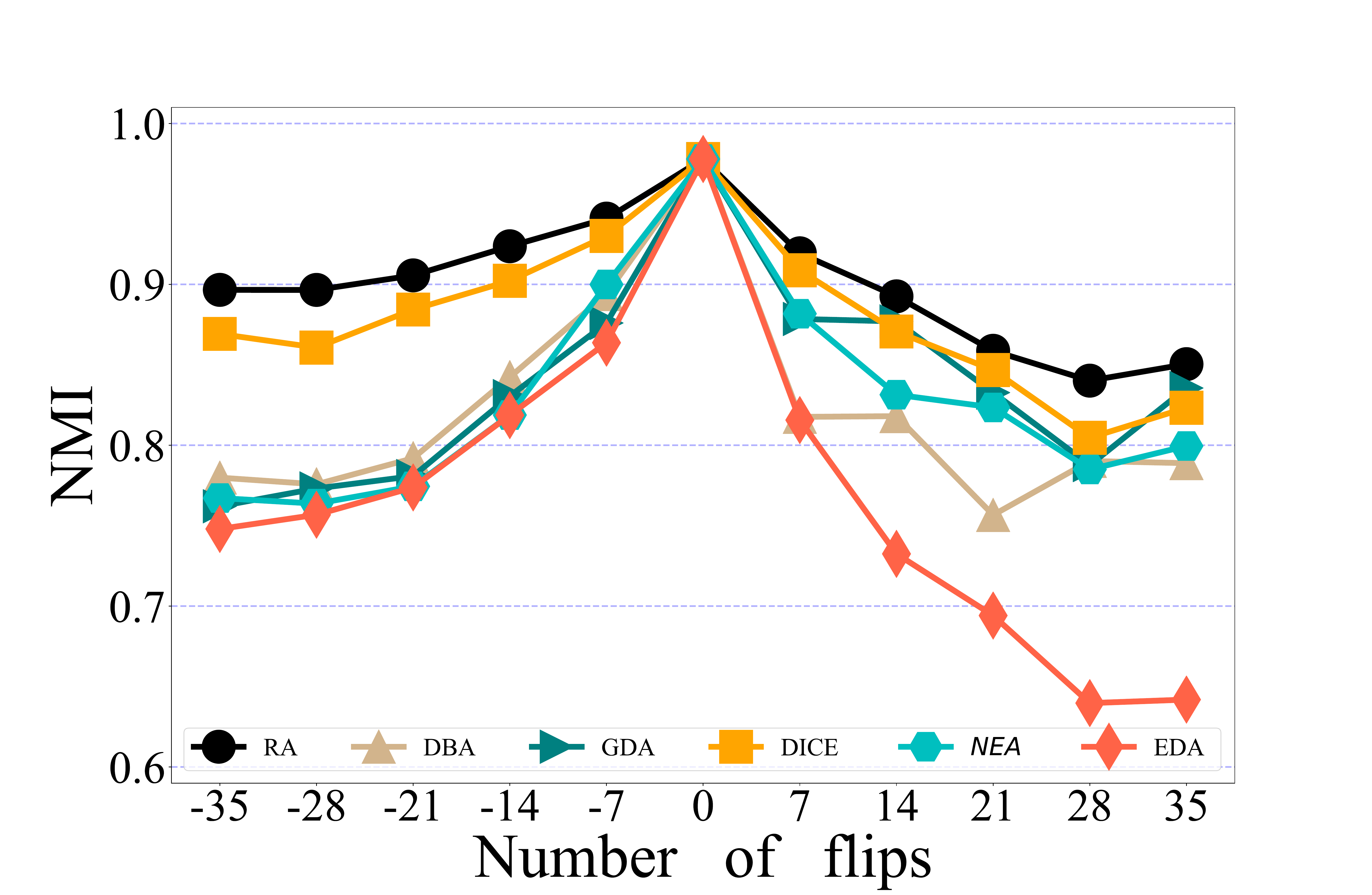}
    }
    \subfigure[citeseer]{
        \includegraphics[width=.23\linewidth]{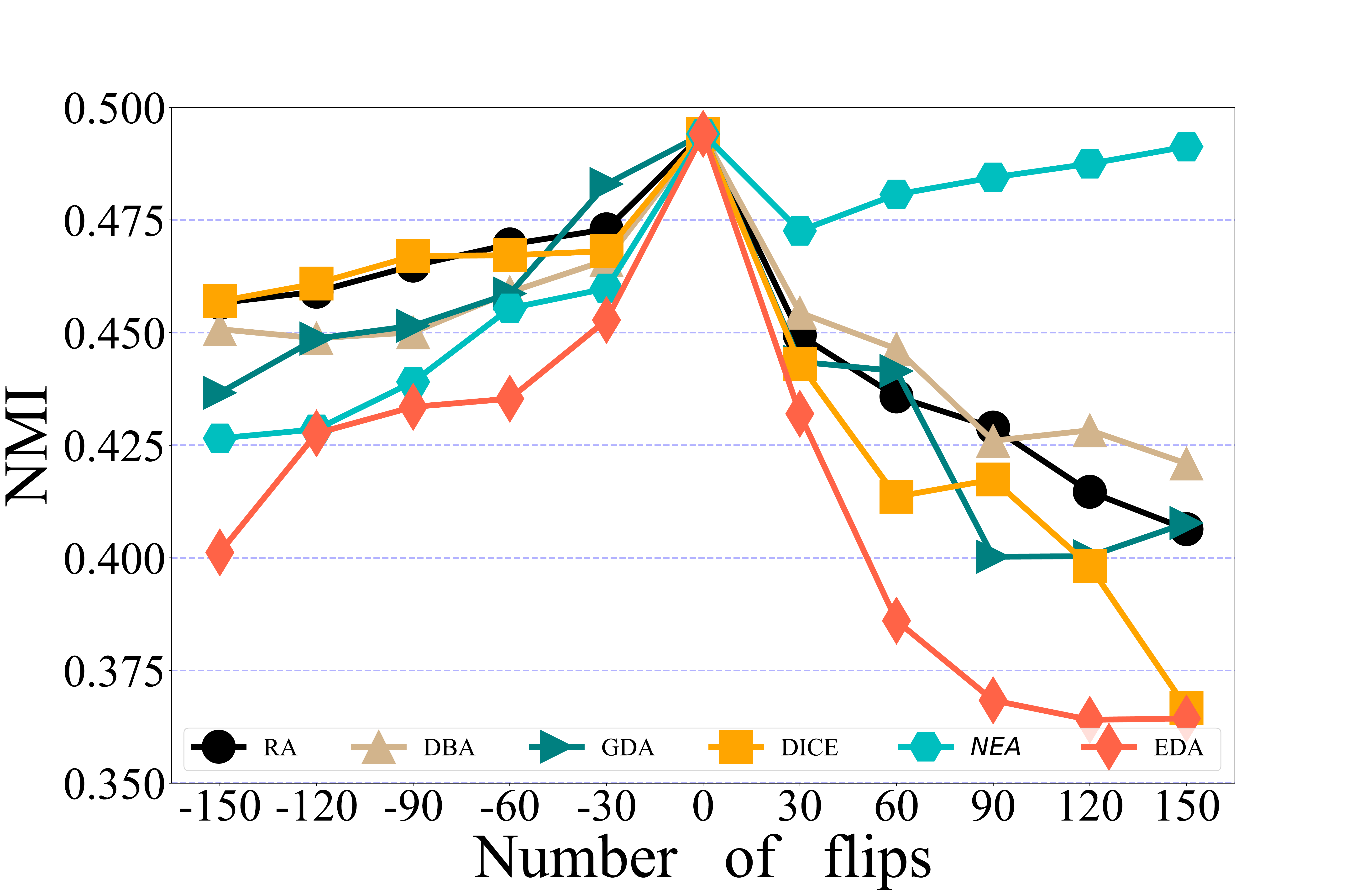}
}
    \subfigure[cora]{
        \includegraphics[width=.23\linewidth]{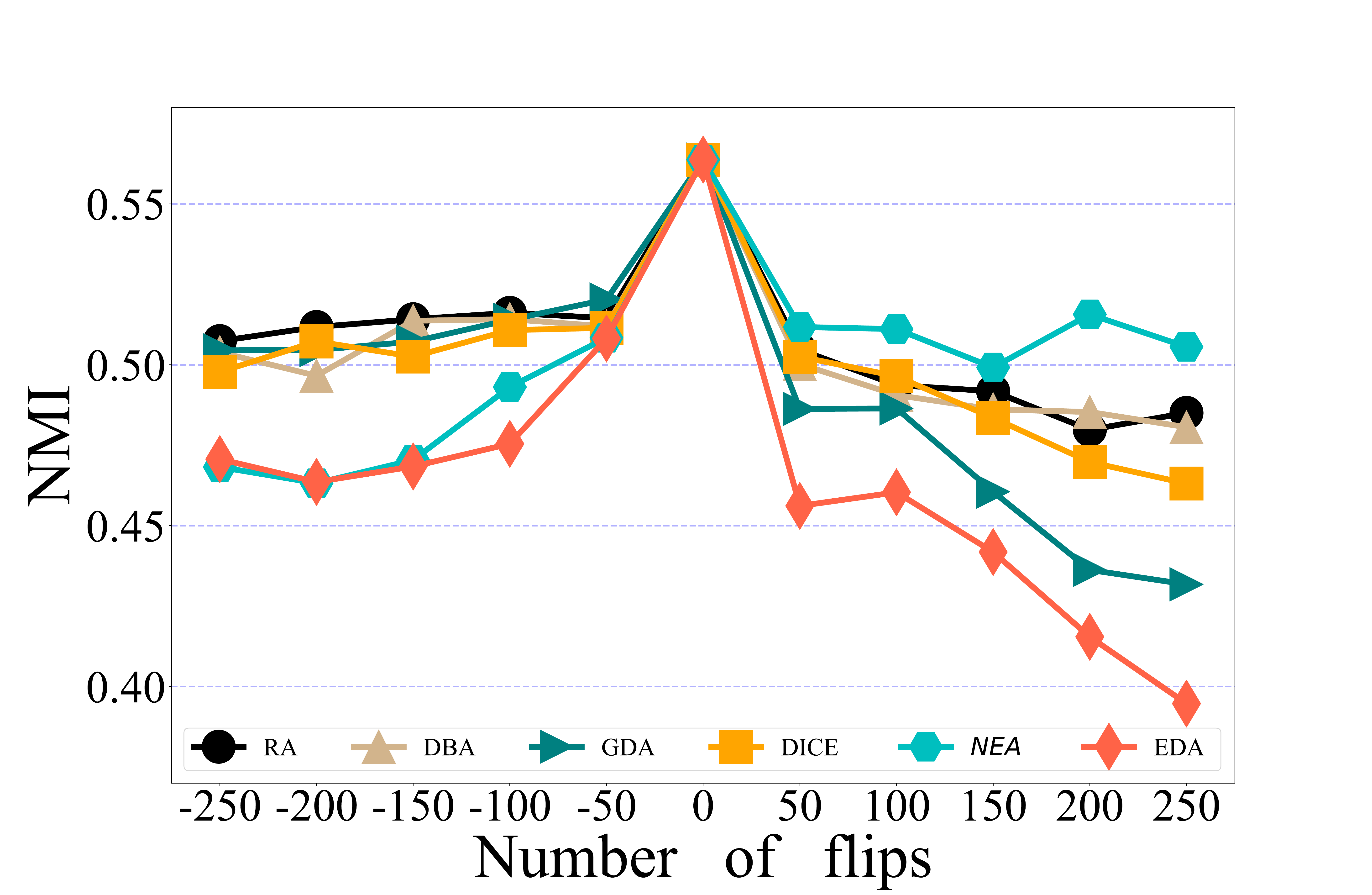}
}
\caption{NMI as the functions of the percentage of attacked links for different attack strategies on community detection.}
\label{Fig:cluster}
\end{figure*}

\begin{figure*}[!t]
\centering
    \subfigure[Karate]{
        \includegraphics[width=.23\linewidth]{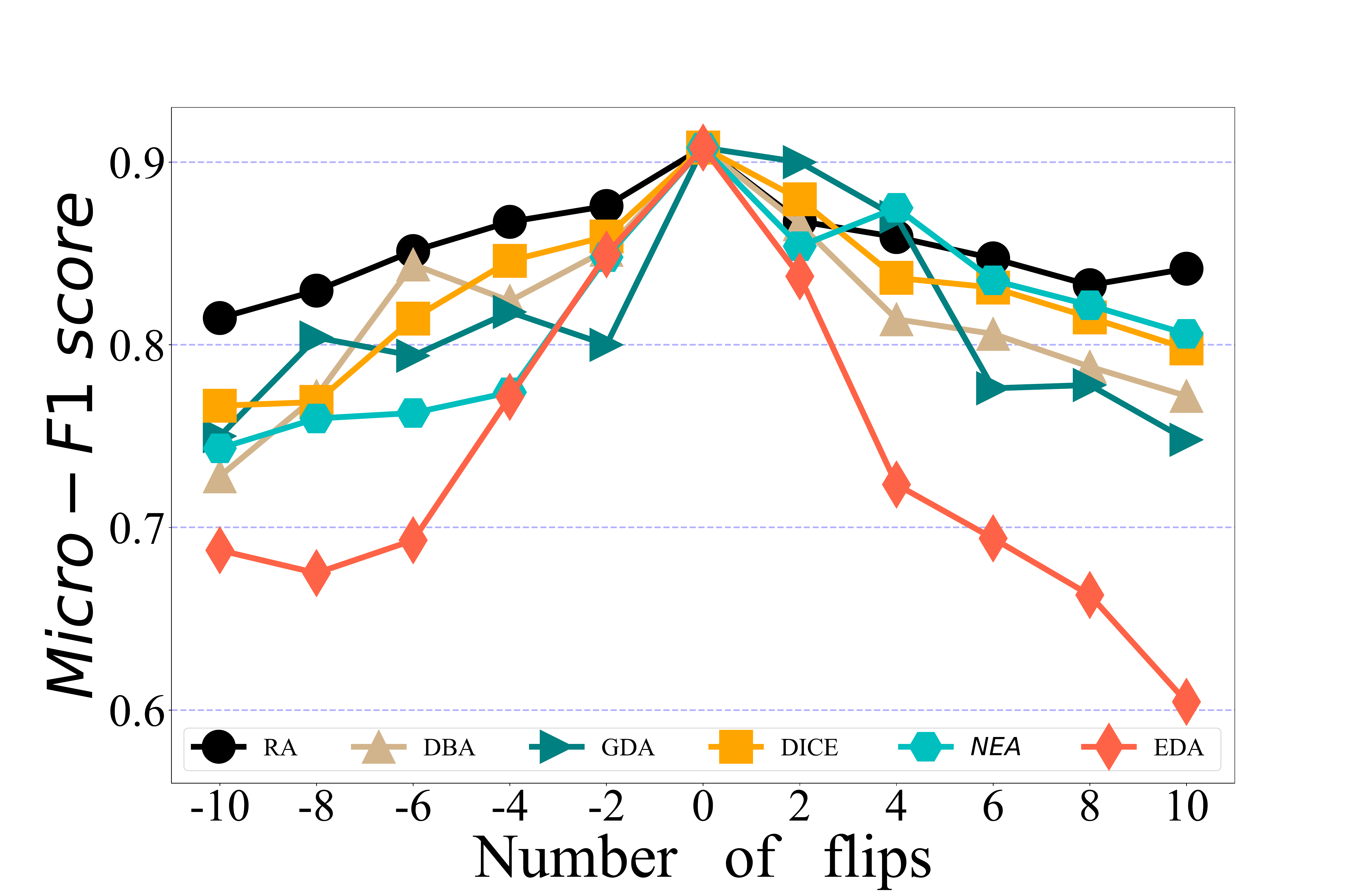}
    }
    \subfigure[Game]{
        \includegraphics[width=.23\linewidth]{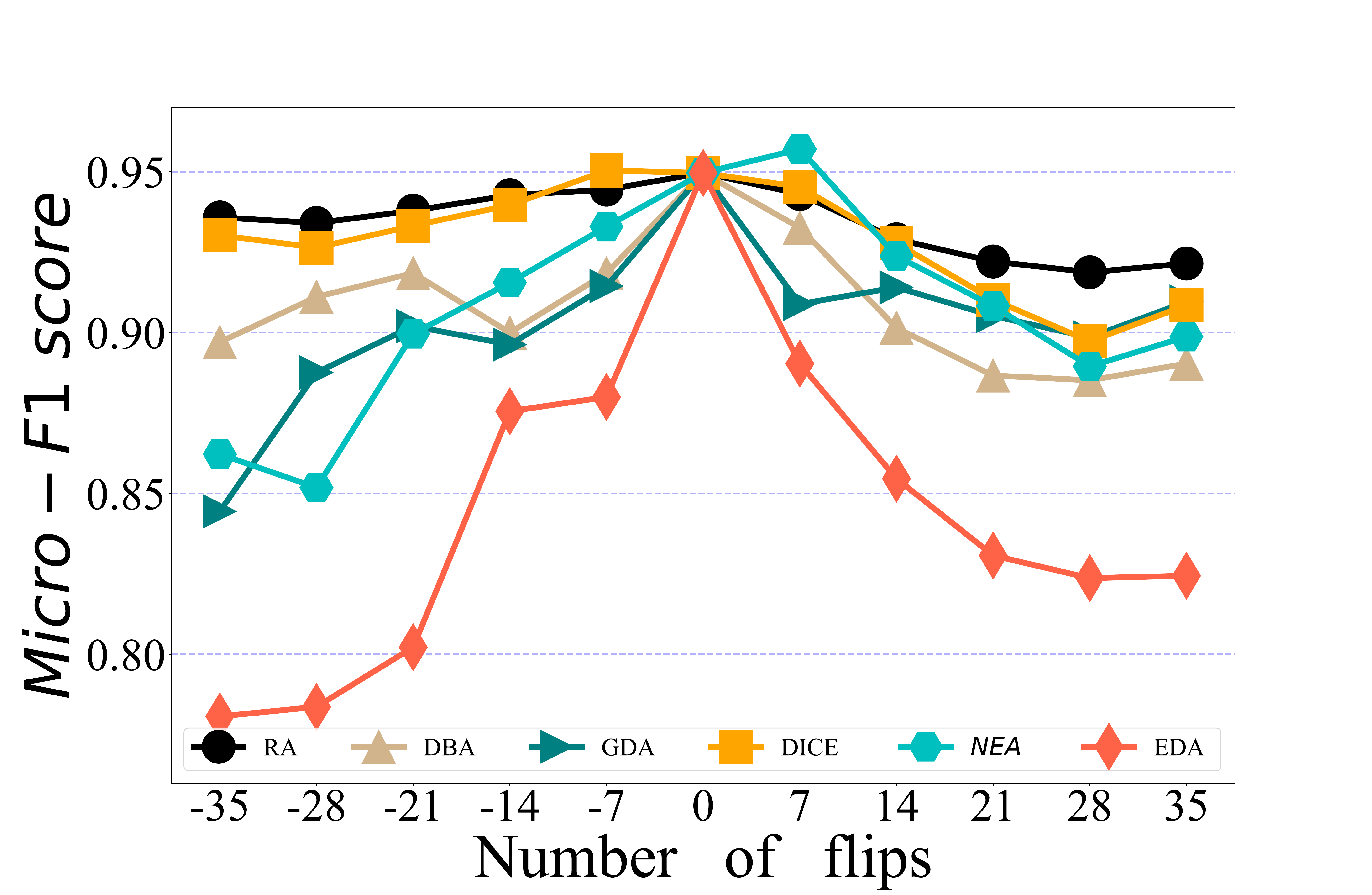}
    }
    \subfigure[citeseer]{
        \includegraphics[width=.23\linewidth]{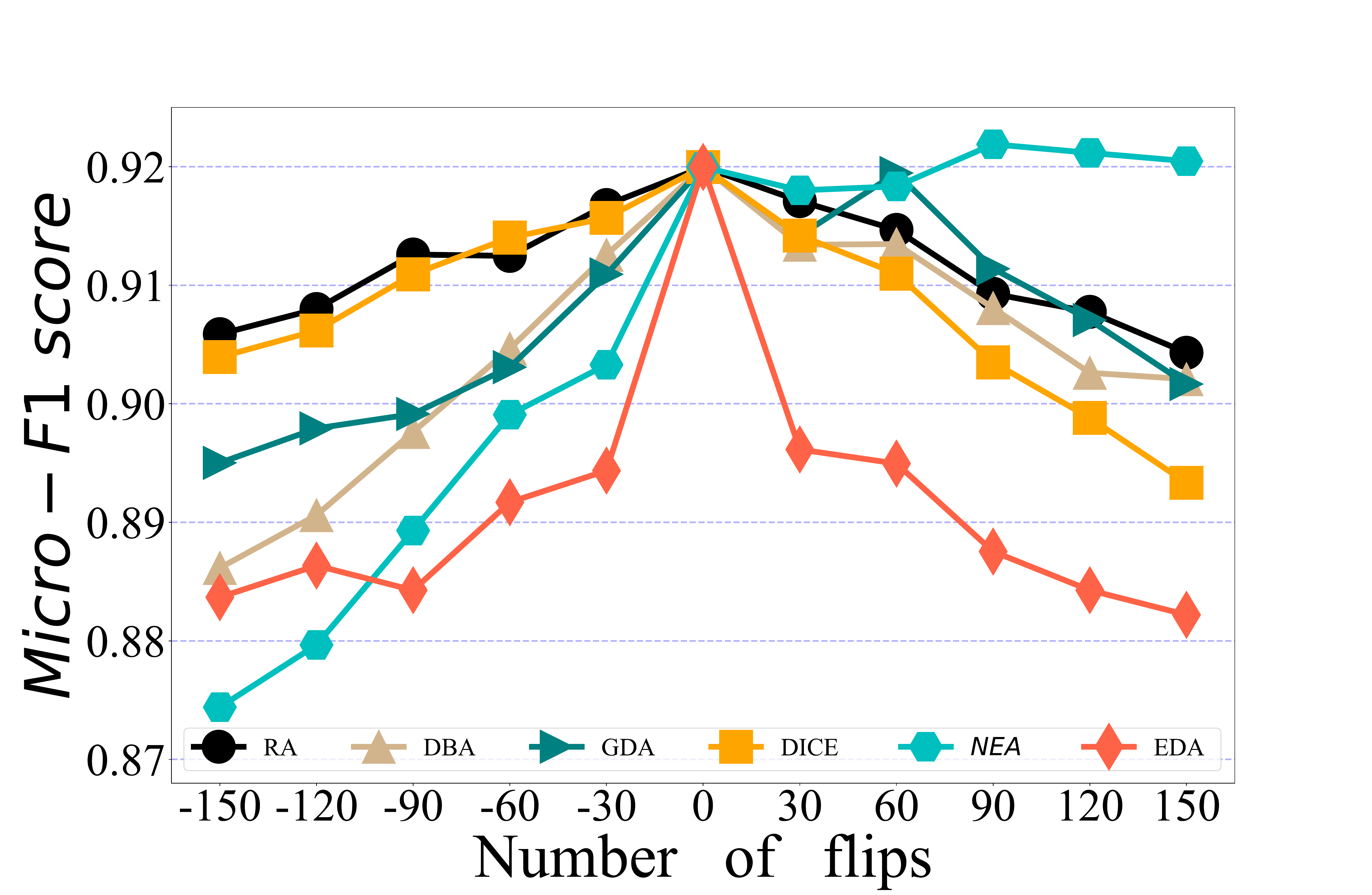}
    }
    \subfigure[cora]{
        \includegraphics[width=.23\linewidth]{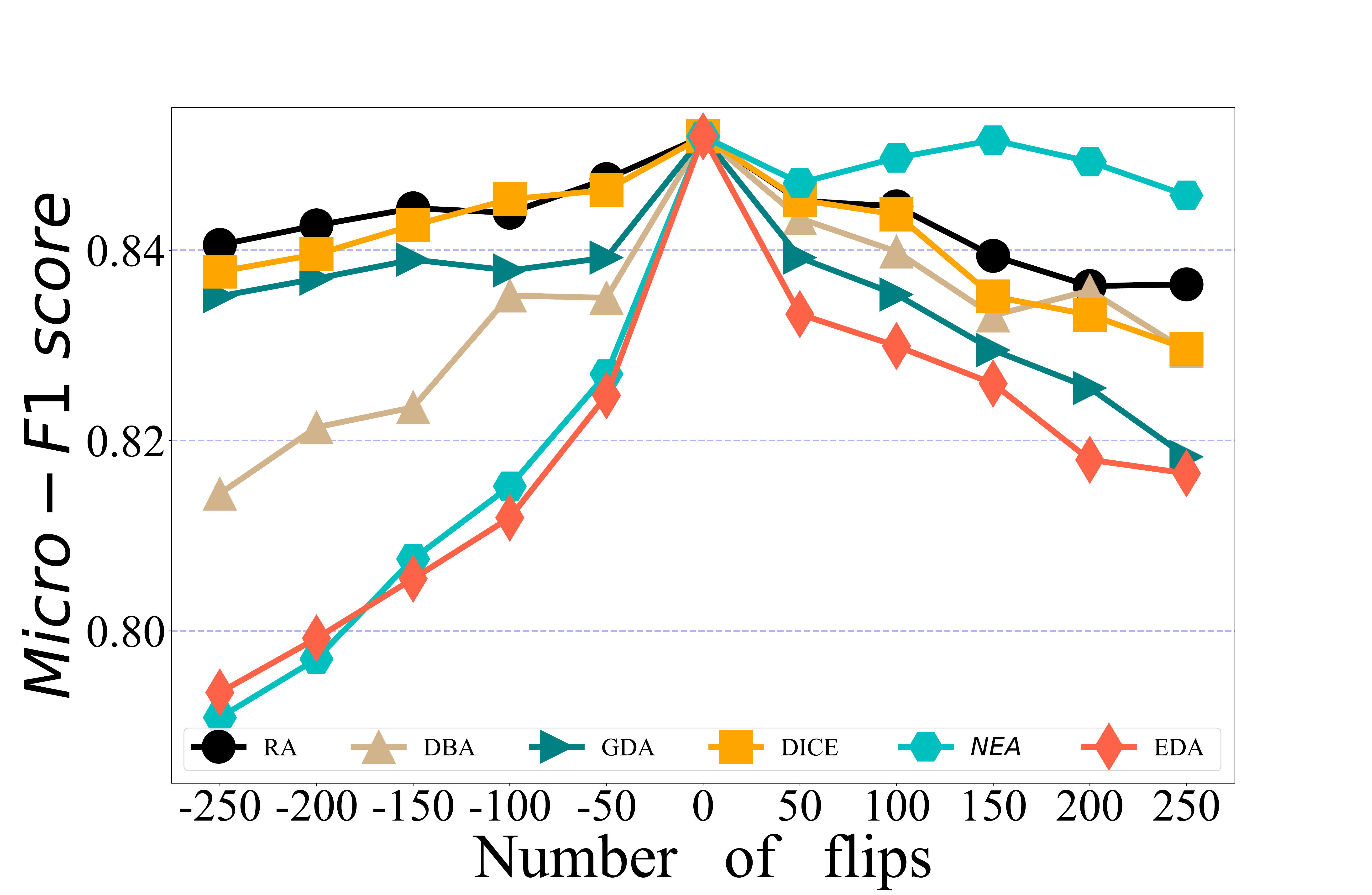}
    }
\caption{Micro-F1 score as the functions of the percentage of attacked links for different attack strategies on node classification.}
 \label{Fig:classificationlr1}
\end{figure*}

\begin{figure*}[!t]
\centering
     \subfigure[Karate]{
        \includegraphics[width=.23\linewidth]{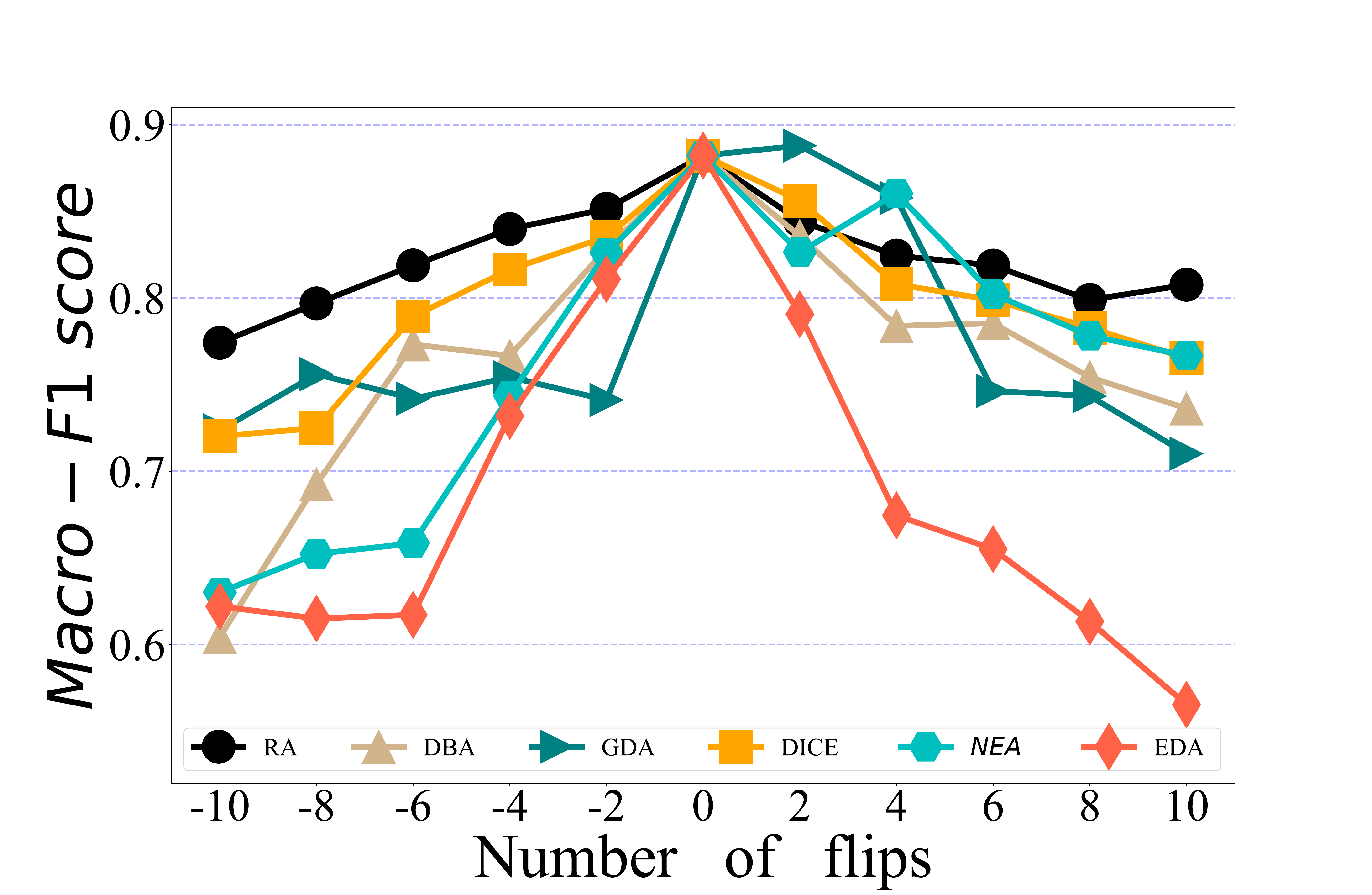}
    }
        \subfigure[Game]{
        \includegraphics[width=.23\linewidth]{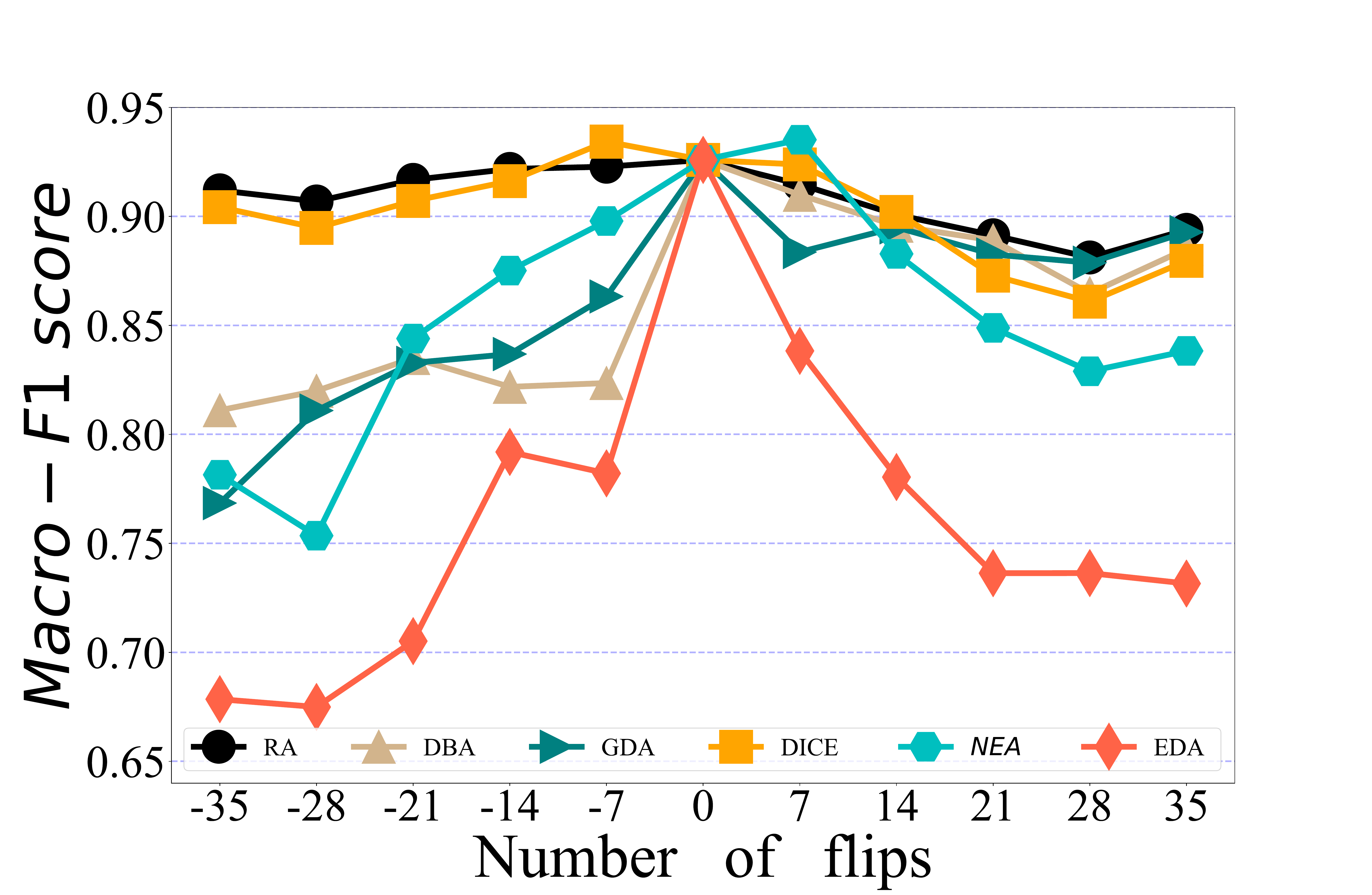}
    }
        \subfigure[citeseer]{
        \includegraphics[width=.23\linewidth]{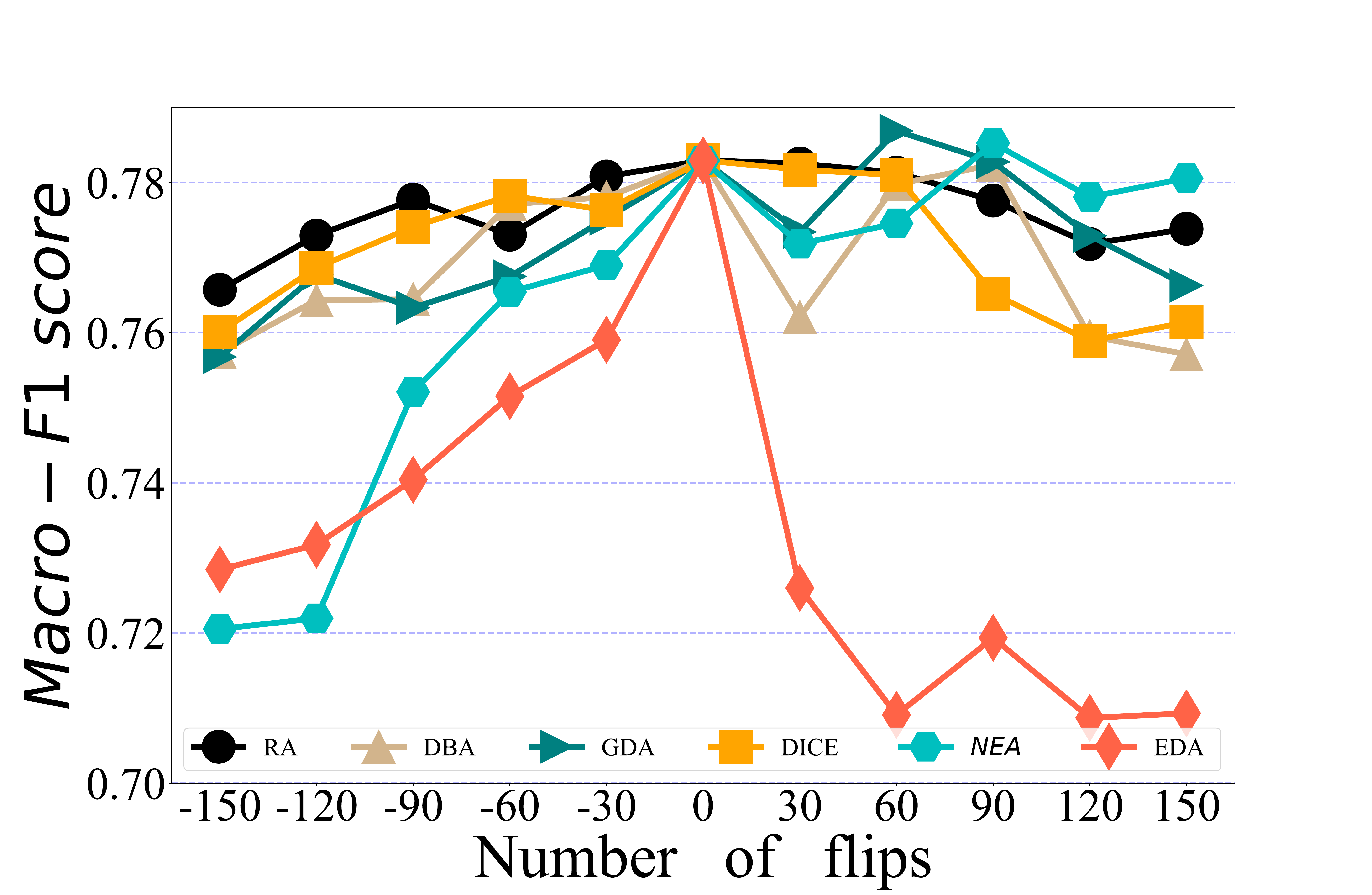}
    }
        \subfigure[cora]{
        \includegraphics[width=.23\linewidth]{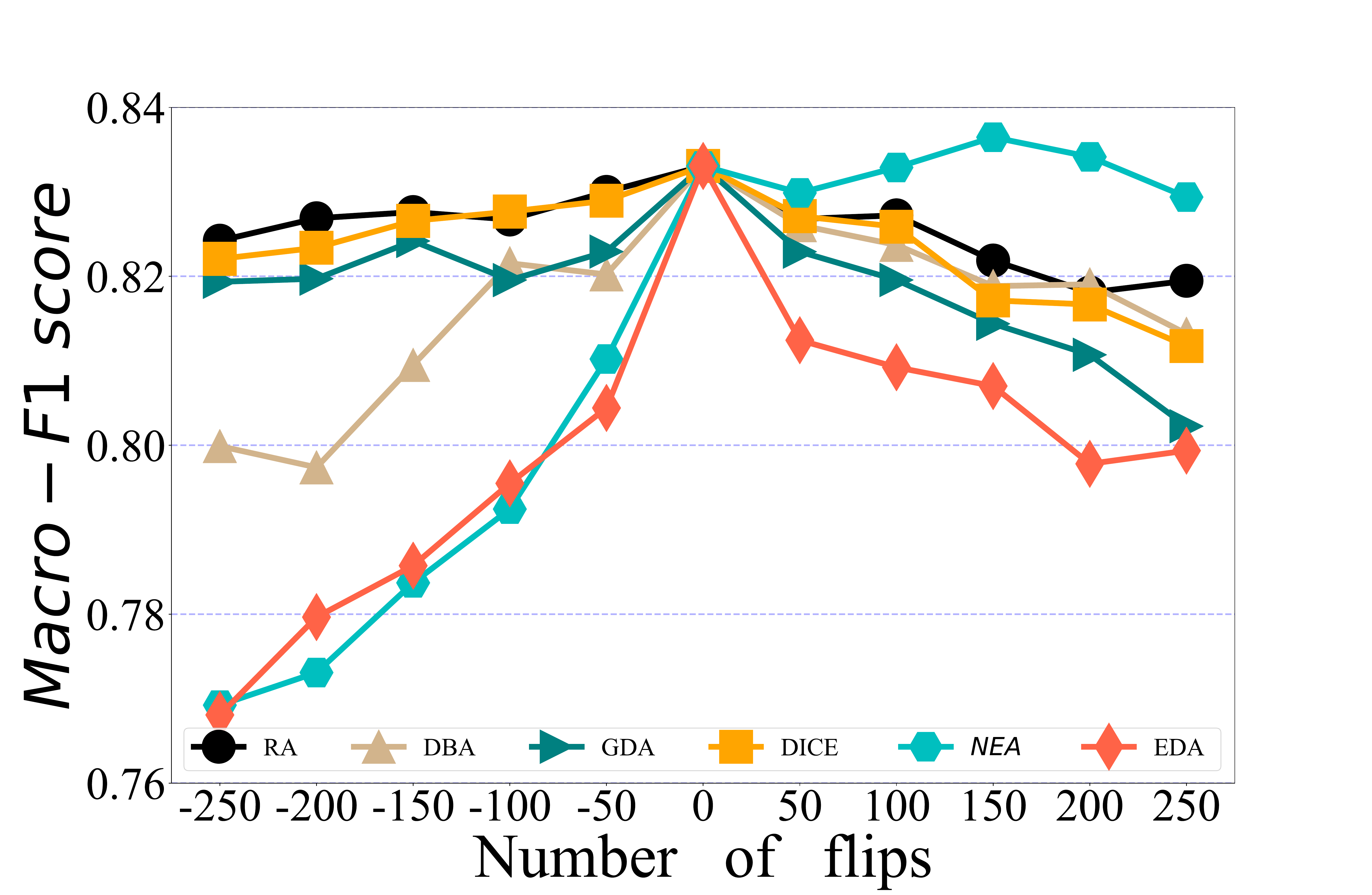}
    }
\caption{Macro-F1 score as the functions of the percentage of attacked links for different attack strategies on node classification.}
 \label{Fig:classificationlr2}
\end{figure*}

\subsection{Attack on community detection}
\label{sec:Community detection algorithm}
Community detection is one of the most common unsupervised learning problems in network science, aiming to identify the communities (a group of nodes that are closely connected with each other) in a whole network. There are many community detection algorithms. Here, to validate the effectiveness of different attack strategies on network embedding, we would prefer to transfer nodes into vectors by DeepWalk and then realize the community detection by clustering these vectors in the embedding space by using K-means algorithm.

For each attack strategy, we flip the same links and then use the above community detection method to identify communities. We use the Normalized Mutual Information (NMI) to measure the performance~\cite{ghosh2010community}.

NMI is used to evaluate the accuracy of a detected community. For two different categories of prediction $C_p$ and reality $C_t$, it is defined as
\begin{equation}
	{\rm NMI}(C_p,C_t) = \frac{{\rm MI}(C_p,C_t)}{\sqrt{H(C_p)H(C_t)}},
\end{equation}
where MI and $H$ represent the Mutual Information and entropy, respectively, which are defined as

\begin{equation}
	{\rm MI}(C_p,C_t) = \sum_{i=1}^{\left | C_p \right |}\sum_{j=1}^{\left | C_t \right |}P(i,j)log(\frac{P(i,j)}{P(i)P'(j)}),
\label{Eq:MI}
\end{equation}

\begin{equation}
    H(C_p) = \sum_{i=1}^{\left | C_p \right |}P(i)log(P(i)),
\label{Eq:hp}
\end{equation}

\begin{equation}
    H(C_t) = \sum_{j=1}^{\left | C_t \right |}P'(j)log(P'(j)),
\label{Eq:ht}
\end{equation}
respectively, where $|C_p|$, $|C_t|$ are the number of categories in the division of prediction and that of truth, respectively, $P(i) = |C_p^i|/|V|$, $P'(j) = |C_t^j|/|V|$, and $P(i,j) = |C_p^i \cap C_t^j|/|V|$. The value of NMI indicates the similarity between $|C_p|$ and $|C_t|$, thus, the larger value, the more similar the prediction and the truth are.

For baseline attack strategies, we carry out the experiments for $100$ times and present the average result in Fig.~\ref{Fig:cluster}. In each line chart, the highest value in the middle represented is the result without suffering from attack. The left half is the result of deleting links, and the right half is the result of adding links. In general, all proposed attack strategy can effectively reduce the accuracy of community detection. More specifically, heuristic attack strategies, such as DICE and DBA, are more effective than RA.
GDA always has a good effect when there are fewer links fliping.
In some cases, the performance of the NEA may be better than EDA. But in the vast majority of cases, our proposed EDA exhibits the best overall performance, with the lowest $NMI$ in most of the four networks.

\subsection{Attack on node classification}
Different from the community detection problem, node classification is a typical supervised learning in network science in which the label of some nodes are known. Again, here we would like to use DeepWalk to map nodes to vectors and then use Logistic Regression (LR)~\cite{dreiseitl2002logistic} to classify them, namely \emph{DeepWalk+LR}. We use the same set of benchmark networks since their real communities are known beforehand. We randomly choose 80\% of nodes as the training set and treat the rest as the test set, and use Micro-F1 and Macro-F1 to evaluate the classification results.
We calculate the number of true positives ($TP$), false positives ($FP$), true negatives ($TN$), false negatives ($FN$) in the instances.

Macro-F1 and Micro-F1 are then defined as

\begin{equation}
{\rm Macro-F1} = \frac{\sum _{{C_P}\in{C_T}}F1(C_P))}{|C_T|},
\label{Eq:macrof1}
\end{equation}

\begin{equation}
{\rm Micro-F1} = \frac{2*Pr*R}{Pr+R},
\label{Eq:microf1}
\end{equation}
respectively,where $C_P$, $C_T$ are the category of in the division of prediction and that of truth, $F1(C_P)$ is the F1-measure for the label $C_P$, and $Pr$ and $R$ are calculated by
\begin{equation}
Pr = \frac{\sum _{{C_P} \in {C_T}}TP}{\sum _{{C_P} \in {C_T}}(TP+FP)},
\label{Eq:Pr}
\end{equation}

\begin{equation}
R = \frac{\sum _{{C_P} \in {C_T}}TP}{\sum _{{C_P} \in {C_T}}(TP+FN)},
\label{Eq:R}
\end{equation}
respectively. For multi-classification problems, Micro-F1 and Macro-F1 are used to evaluate the performance of the classification model.

 \begin{figure*}[t]
 \centering
    \includegraphics[width=0.9\linewidth]{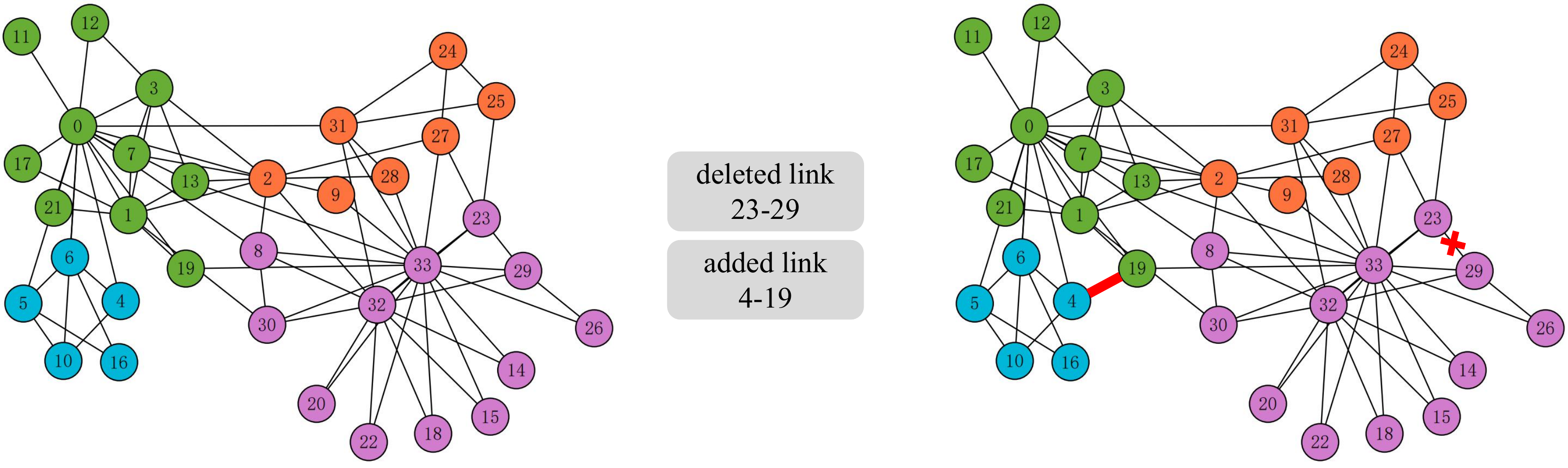}
        \caption{The network visualization of EDA attack. Left is the original karate network, and the right is the adversarial network generated by EDA, which added a link between nodes $4$ and $19$, while deleted a link between nodes $23$ and $29$. Different colors represent different communities.}
        \label{Fig:net-attack}
 \end{figure*}

Similarly, for each attack strategy, we flip a few links and then use the above two indicators to evaluate. For each case, we carry out the experiments for 100 times and present the average of results in Fig.~\ref{Fig:classificationlr1} and Fig.~\ref{Fig:classificationlr2}. We find that both Micro-F1 and Macro-F1 decrease after each attack, regardless of the choice of the downstream classification algorithm (LR or KNN). In most cases, EDA still performs best, leading to the biggest drop of Micro-F1 and Macro-F1.
The heuristic attack strategies DICE is more effective than RA, consistent with the results in community detection. However, it seems that in some instances, when the percentage of flipping links is relatively big, baseline methods may be more effective than EDA.
It might be due to:
\begin{itemize}
    \item First, comparing with DICE, which is a white-box attack method based on the real labels of nodes, EDA is a black-box attack method without using label information.
    \item Second, GA has been recognized to tend to be trapped in local optimum. But still, in the vast majority of cases, our proposed EDA is significantly more effective than the other attack strategies.
\end{itemize}

\begin{figure*}[b]
    \centering
        \subfigure[unattack]{
            \includegraphics[width=.23\linewidth]{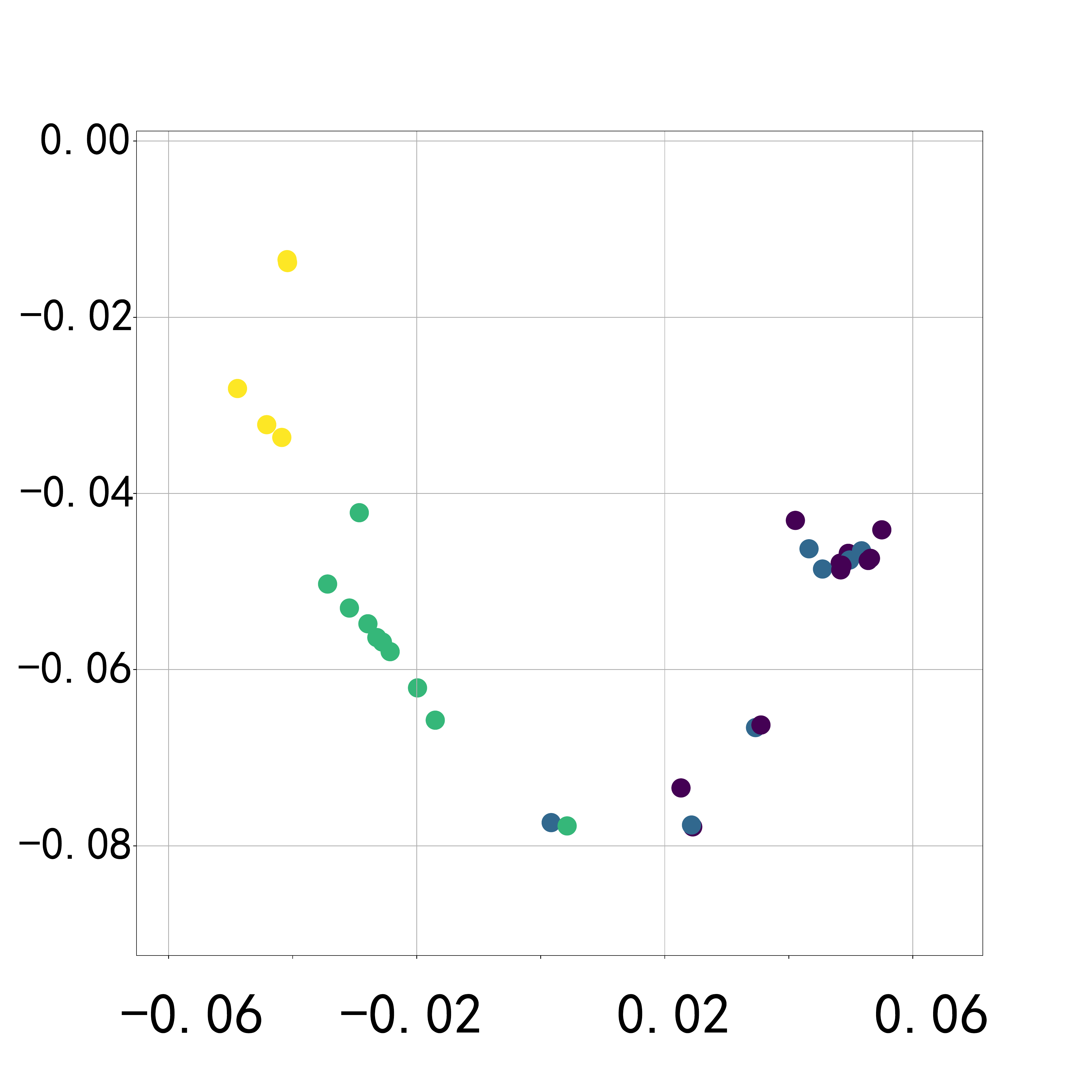}
        }
        \subfigure[attack-1\%]{
            \includegraphics[width=.23\linewidth]{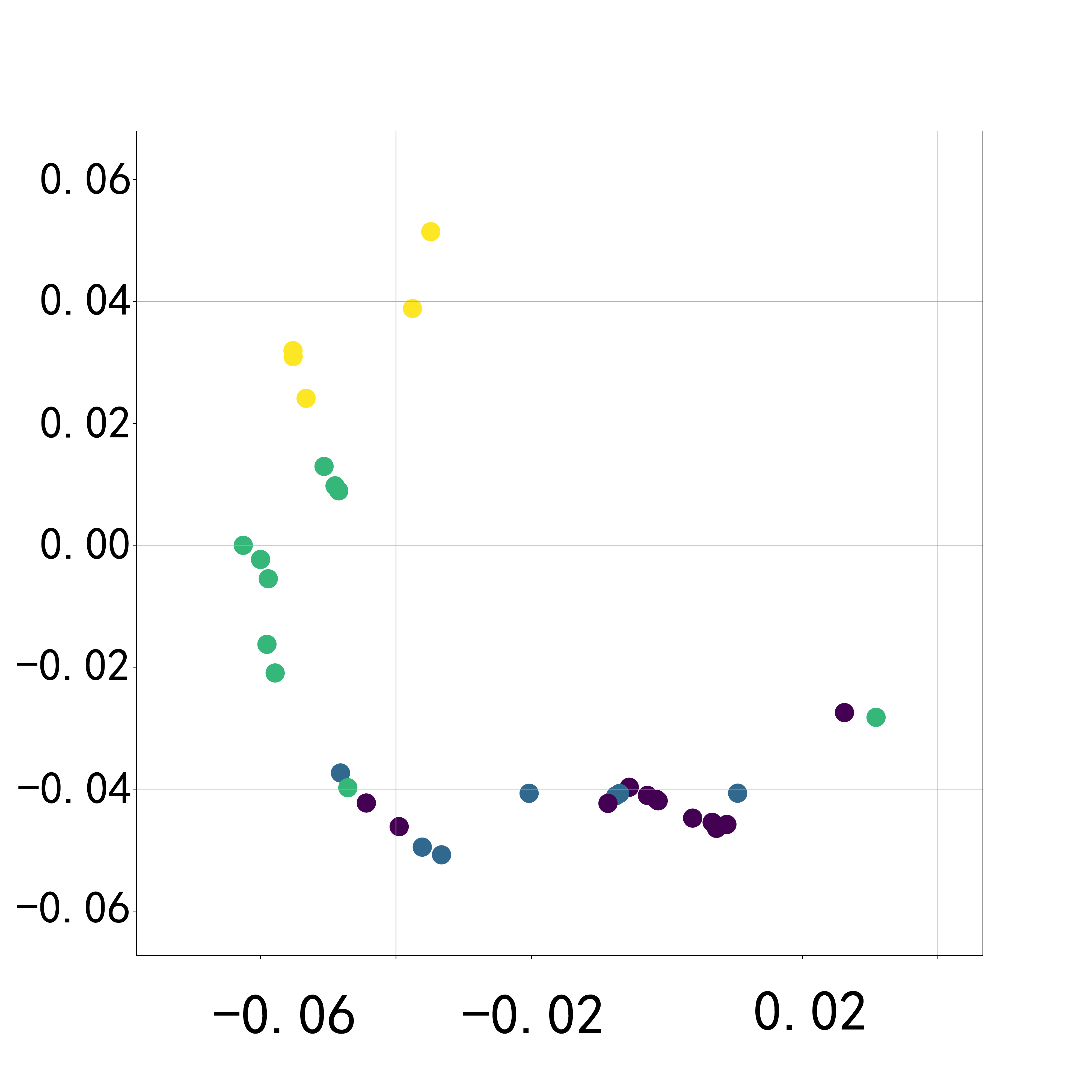}
        }
         \subfigure[attack-2\%]{
            \includegraphics[width=.23\linewidth]{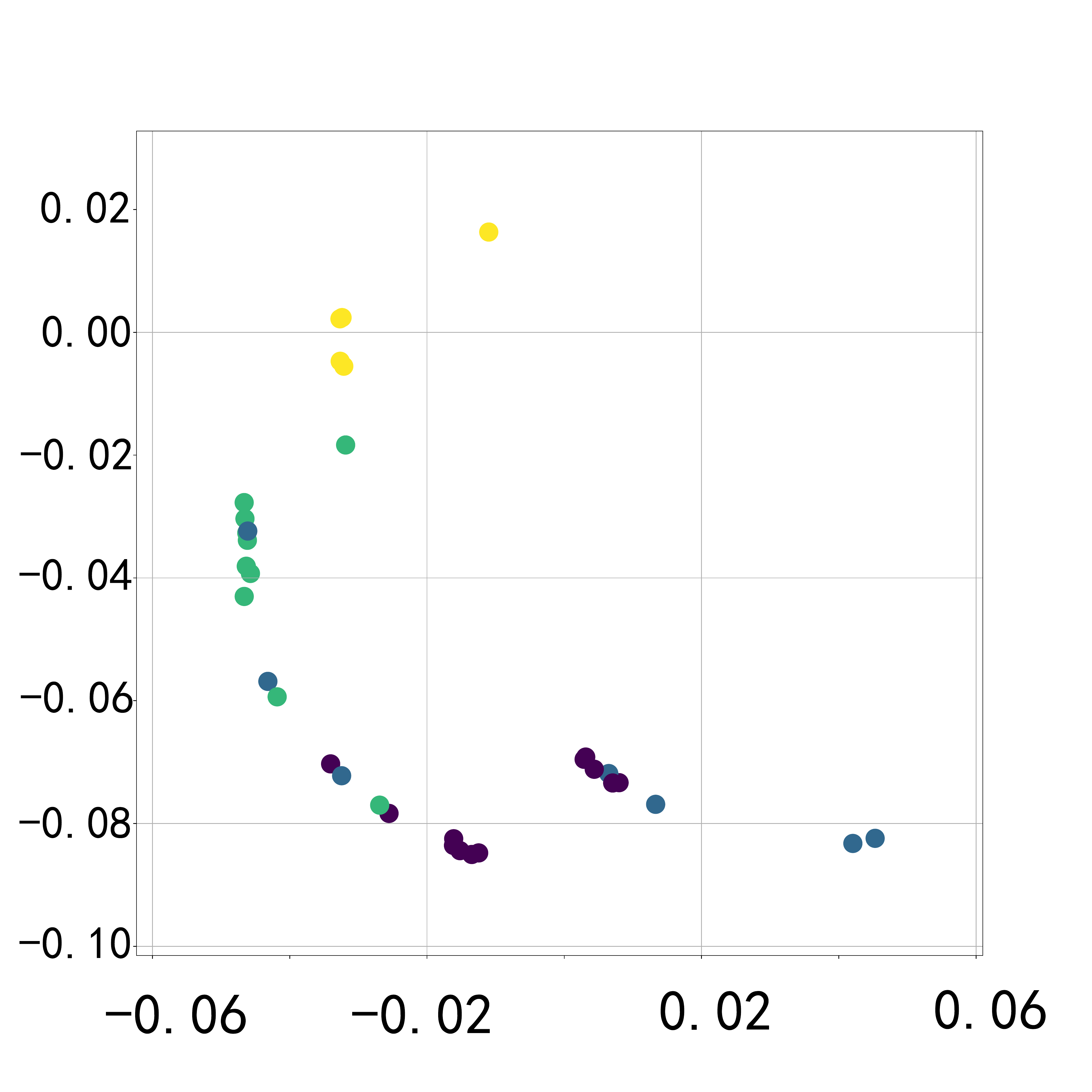}
        }
        \subfigure[attack-3\%]{
            \includegraphics[width=.23\linewidth]{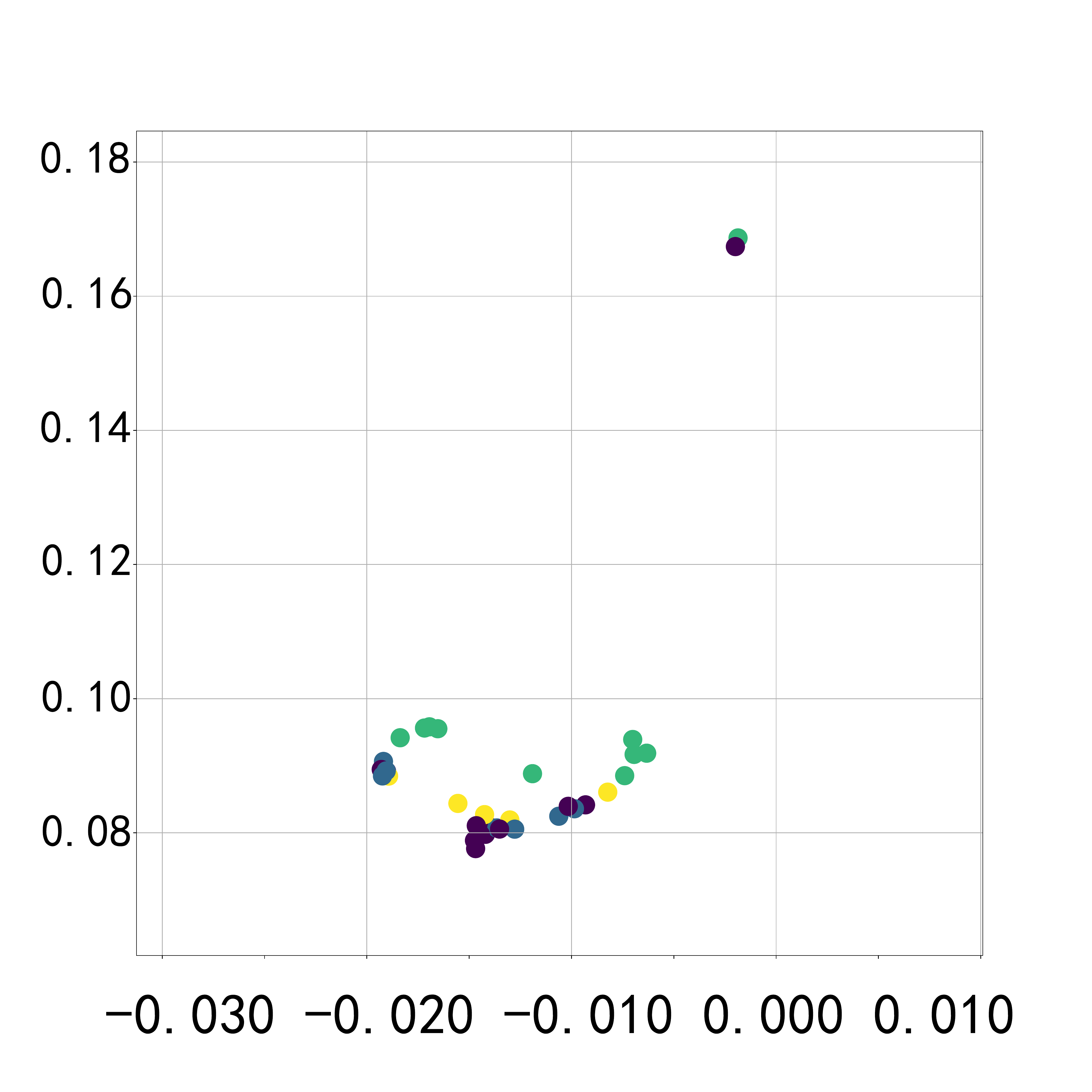}
        }
        \subfigure[attack-4\%]{
            \includegraphics[width=.23\linewidth]{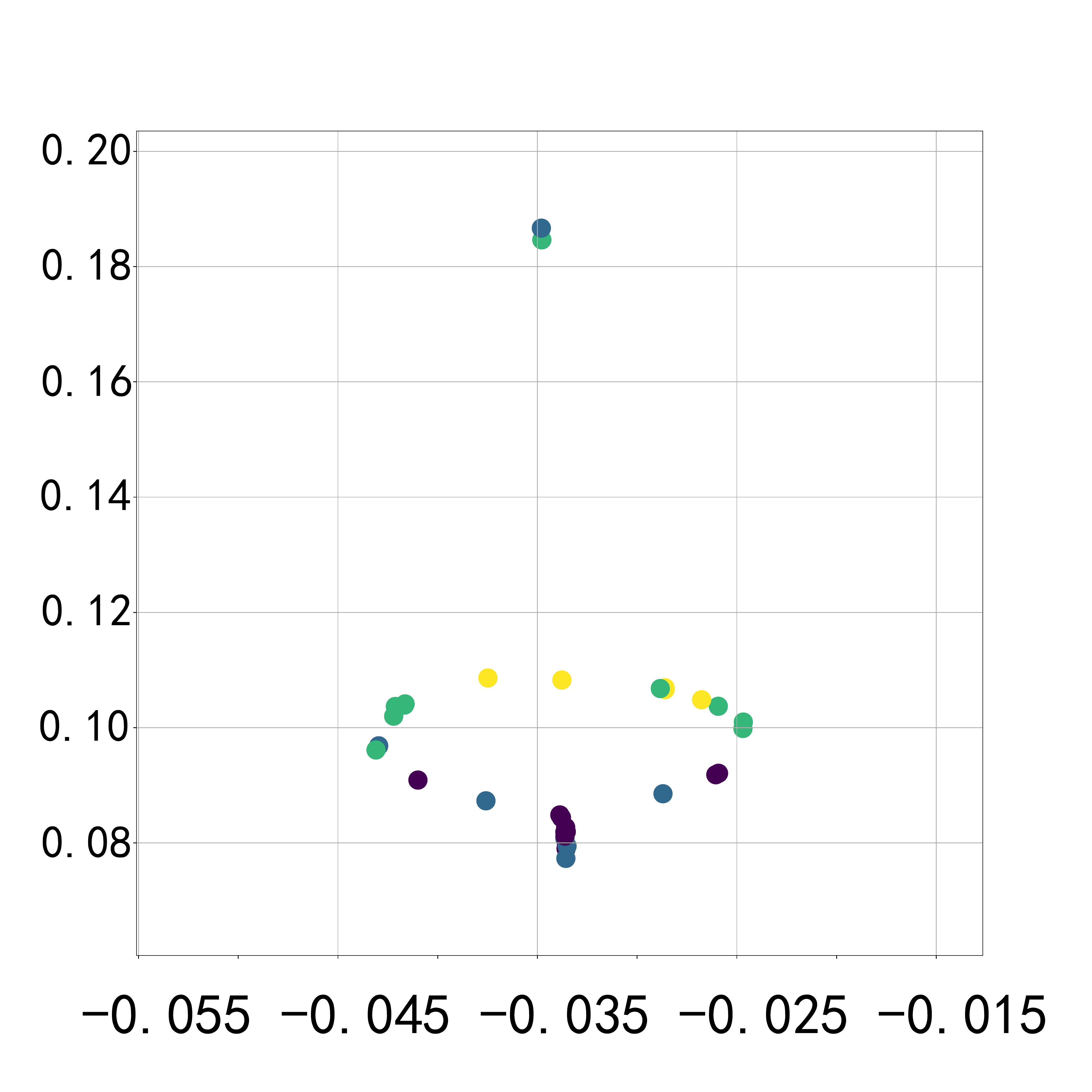}
        }
        \subfigure[attack-5\%]{
            \includegraphics[width=.23\linewidth]{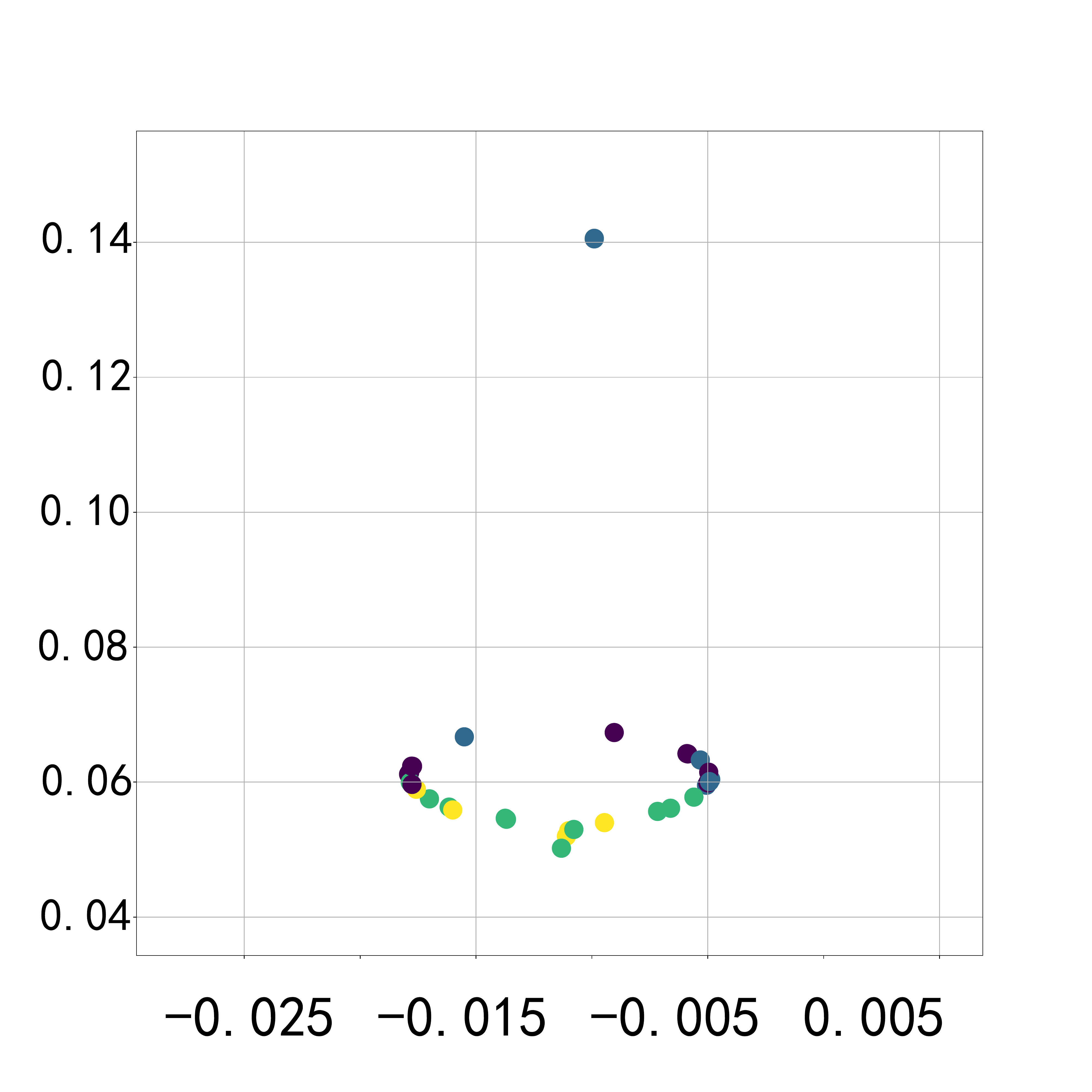}
        }
        \subfigure[attack-6\%]{
            \includegraphics[width=.23\linewidth]{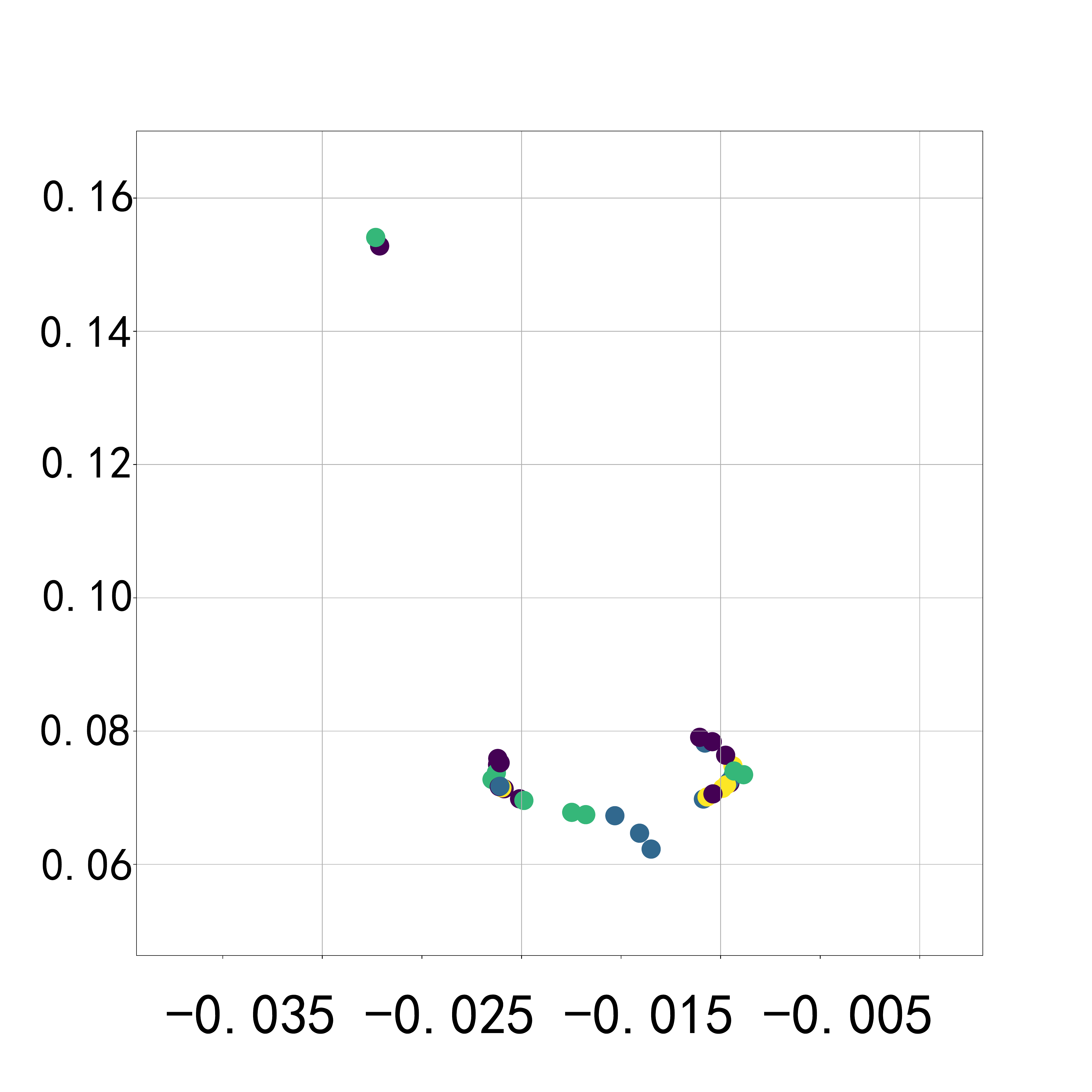}
        }
        \subfigure[attack-7\%]{
            \includegraphics[width=.23\linewidth]{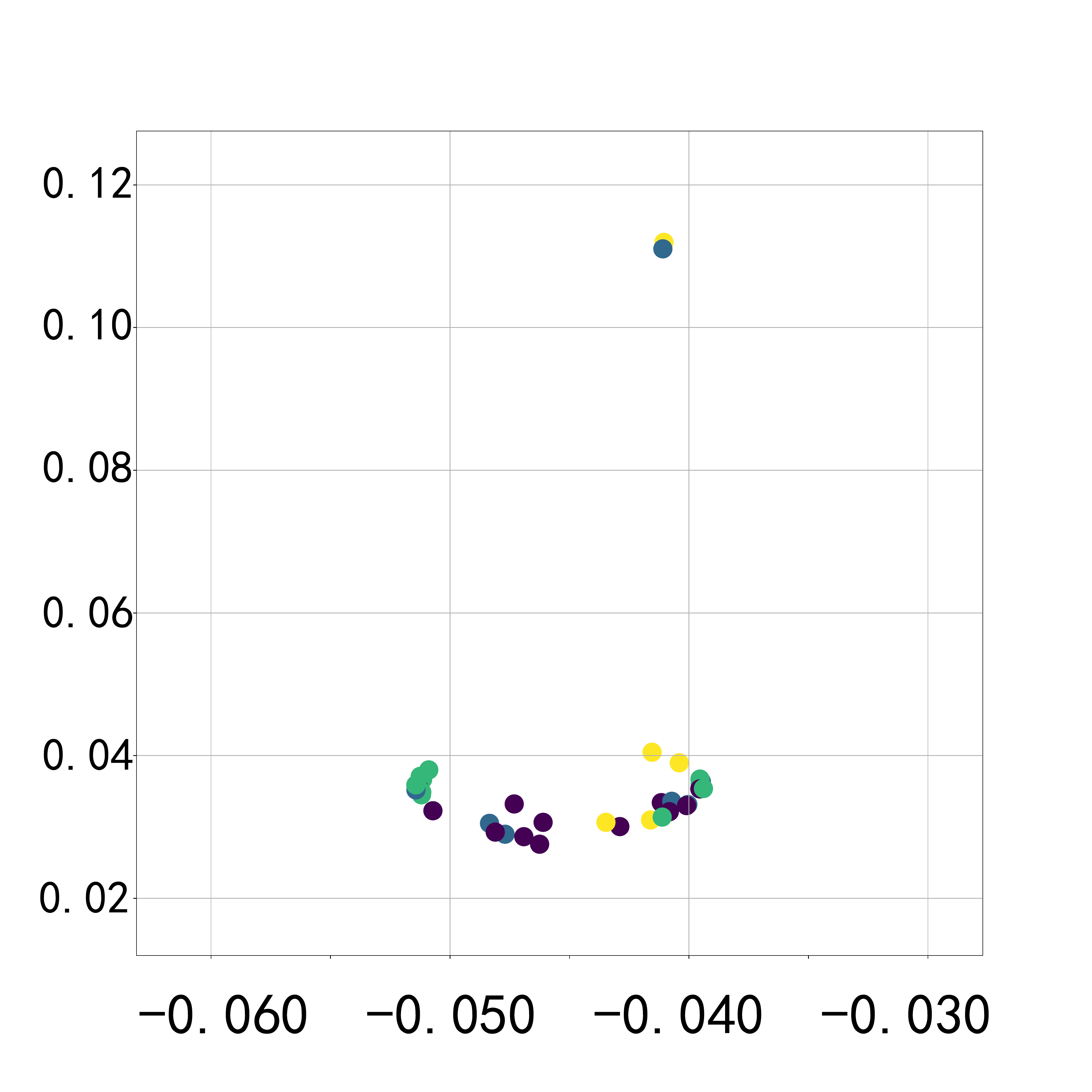}
        }
    \caption{The visualization of node vectors in two-dimensional embedding space by attacking certain percentages of links. Different colors represent different communities. As more links are attacked by EDA, it is getting more difficult to distinguish the node vectors of different communities in the embedding space.}\label{Fig:clustervis}
    \end{figure*}

\begin{table*}[htbp]
  \centering
  \caption{Transferability results on different attack strategies.}
  \setlength{\tabcolsep}{3.3mm}
  \renewcommand\arraystretch{1.1}

    \begin{tabular}{c|c|c|c|cccccc}
    \toprule
    \hline
    Dataset & model & Metric  & Network & \multicolumn{1}{c}{1\%} & \multicolumn{1}{c}{2\%} & \multicolumn{1}{c}{3\%} & \multicolumn{1}{c}{4\%} & \multicolumn{1}{c}{5\%} & 6\% \\
      \hline

    \multirow{30}[15]{*}{Game} & \multirow{12}[6]{*}{HOPE+LR} & \multirow{6}[3]{*}{Micro-F1} & Unattack & 0.9133  & 0.9133  & 0.9133  & 0.9133  & 0.9133  & 0.9133  \\
          &       &       & RA    & 0.8966  & 0.8868  & 0.8774  & 0.8680  & 0.8599  & 0.8539  \\
          &       &       & RLS    & 0.8963&	0.8923&	0.889&	0.8857&	0.8814&	0.8769  \\
          &       &       & DBA    & 0.8982&	0.8940&	0.8904&	0.8867&	0.8827&	0.8790   \\
          &       &       & DICE  & 0.8923  & 0.8799  & 0.8679  & 0.8545  & \textbf{0.8387 } & \textbf{0.8291} \\
          &       &       & EDA*  & \textbf{0.8659} & \textbf{0.8594} & \textbf{0.8516} & \textbf{0.8434} & 0.8394  & 0.8365  \\
\cline{3-10}          &       & \multirow{6}[3]{*}{Macro-F1} & Unattack & 0.8688  & 0.8688  & 0.8688  & 0.8688  & 0.8688  & 0.8688  \\
          &       &       & RA    & 0.8520  & 0.8393  & 0.8247  & 0.8116  & 0.8013  & 0.7596  \\
          &       &       & RLS    & 0.8504&	0.8452&	0.8414&	0.8356&	0.8293&	0.8230  \\
          &       &       & DBA    & 0.8553&	0.8496&	0.8445&	0.8401&	0.8350&	0.8299  \\
          &       &       & DICE  & 0.8462  & 0.8290  & 0.8125  & 0.7951  & 0.7734  & \textbf{0.7197 } \\
          &       &       & EDA*  & \textbf{0.8098} & \textbf{0.7997} & \textbf{0.7876} & \textbf{0.7750} & \textbf{0.7683} & 0.7659  \\
\cline{2-10}          & \multirow{6}[3]{*}{HOPE+K-Means} & \multirow{6}[3]{*}{NMI} & Unattack & 0.4832  & 0.4832  & 0.4832  & 0.4832  & 0.4832  & 0.4832  \\
          &       &       & RA    & 0.4767  & 0.4757  & 0.4731  & 0.4699  & 0.4674  & 0.4621  \\
          &       &       & RLS    & 0.4753&	0.4717&	0.4672&	0.4594&	0.4579&	0.4494   \\
          &       &       & DBA    & 0.7868  & 0.7744  & 0.7570  & 0.7444  & 0.7242  & 0.7107  \\
          &       &       & DICE  & 0.4740  & 0.4706  & 0.4627  & 0.4590  & 0.4514  & \textbf{0.4464 } \\
          &       &       & EDA*  & \textbf{0.4720} & \textbf{0.4669} & \textbf{0.4606} & \textbf{0.4572} & \textbf{0.4512} & 0.4475  \\
\cline{2-10}          & \multirow{6}[3]{*}{LPA} & \multirow{6}[3]{*}{NMI} & Unattack & 0.6717  & 0.6717  & 0.6717  & 0.6717  & 0.6717  & 0.6717  \\
          &       &       & RA    & 0.6554  & 0.6349  & 0.6195  & 0.5984  & 0.5837  & 0.5653  \\
          &       &       & RLS    & 0.6563&	0.6442&	0.6256&	0.6098&	0.5979&	0.5745  \\
          &       &       & DBA    & 0.6518&	0.6268&	0.6124&	0.5936&	0.5707&	0.5498  \\
          &       &       & DICE  & 0.6473  & 0.6172  & 0.5909  & 0.5689  & 0.5349  & 0.5105  \\
          &       &       & EDA*  & \textbf{0.6228} & \textbf{0.5905} & \textbf{0.5737} & \textbf{0.5391} & \textbf{0.5262} & \textbf{0.4969} \\
\cline{2-10}          & \multirow{6}[3]{*}{EM} & \multirow{6}[3]{*}{NMI} & Unattack & 0.7233  & 0.7233  & 0.7233  & 0.7233  & 0.7233  & 0.7233  \\
          &       &       & RA    & 0.7174  & 0.7041  & 0.6995  & 0.6911  & 0.6835  & 0.6761  \\
          &       &       & RLS    & 0.7171&	0.7038&	0.7060&	0.6934&	0.6882&	0.6695   \\
          &       &       & DBA    & 0.7170&	0.7099&	0.7022&	0.7000&	0.6921&	0.6901   \\
          &       &       & DICE  & 0.7190  & 0.6999  & 0.6886  & 0.6731  & \textbf{0.6624 } & 0.6585  \\
          &       &       & EDA*  & \textbf{0.6893} & \textbf{0.6860} & \textbf{0.6738} & \textbf{0.6557} & 0.6658  & \textbf{0.6418} \\
      \hline
      \hline
    \multirow{30}[15]{*}{Dolphin} & \multirow{12}[6]{*}{HOPE+LR} & \multirow{6}[3]{*}{Micro-F1} & Unattack & 0.9149  & 0.9149  & 0.9149  & 0.9149  & 0.9149  & 0.9149  \\
          &       &       & RA    & 0.9115  & 0.9081  & 0.9071  & 0.9030  & 0.9000  & 0.8957  \\
          &       &       & RLS    & 0.9102&	0.9089&	0.9078&	0.9063&	0.9046&	0.9032  \\
          &       &       & DBA    & 0.9111&	0.9093&	0.9094&	0.9093&	0.9082&	0.9080  \\
          &       &       & DICE  & 0.9048  & 0.8998  & 0.8896  & 0.8867  & 0.8747  & \textbf{0.8668 } \\
          &       &       & EDA*  & \textbf{0.8861} & \textbf{0.8933} & \textbf{0.8845} & \textbf{0.8798} & \textbf{0.8731} & 0.8691  \\
\cline{3-10}          &       & \multirow{6}[3]{*}{Macro-F1} & Unattack & 0.8943  & 0.8943  & 0.8943  & 0.8943  & 0.8943  & 0.8943  \\
          &       &       & RA    & 0.8926  & 0.8885  & 0.8877  & 0.8825  & 0.8793  & 0.8740  \\
          &       &       & RLS    & 0.8906&	0.8896&	0.8887&	0.8869&	0.8847&	0.8831   \\
          &       &       & DBA    & 0.8919&	0.8896&	0.8896&	0.8900&	0.8884&	0.8882   \\
          &       &       & DICE  & 0.8844  & 0.8781  & 0.8666  & 0.8631  & 0.8483  & \textbf{0.8394 } \\
          &       &       & EDA*  & \textbf{0.8640} & \textbf{0.8710} & \textbf{0.8596} & \textbf{0.8539} & \textbf{0.8458} & 0.8403  \\
\cline{2-10}          & \multirow{6}[3]{*}{HOPE+K-Means} & \multirow{6}[3]{*}{NMI} & Unattack & 0.3661  & 0.3661  & 0.3661  & 0.3661  & 0.3661  & 0.3661  \\
          &       &       & RA    & 0.3651  & 0.3717  & 0.3616  & 0.3732  & 0.3729  & 0.3707  \\
          &       &       & RLS    & 0.3415&	0.3590&	0.3522&	0.3532&	0.3453&	0.3422   \\
          &       &       & DBA    & 0.3480&	0.3383&	0.3395&	0.3431&	0.3460&	0.3455   \\
          &       &       & DICE  & 0.3351  & 0.3335  & 0.3328  & 0.3248  & 0.3226  & 0.3142  \\
          &       &       & EDA*  & \textbf{0.3301} & \textbf{0.3204} & \textbf{0.3183} & \textbf{0.3210} & \textbf{0.3167} & \textbf{0.2964} \\
\cline{2-10}          & \multirow{6}[3]{*}{LPA} & \multirow{6}[3]{*}{NMI} & Unattack & 0.6751  & 0.6751  & 0.6751  & 0.6751  & 0.6751  & 0.6751  \\
          &       &       & RA    & 0.6569  & 0.6513  & 0.6406  & 0.6268  & 0.6217  & 0.6143  \\
          &       &       & RLS    & 0.6622&	0.6585&	0.6469&	0.6406&	0.6279&	0.6305   \\
          &       &       & DBA    & 0.6490&	0.6375&	0.6207&	0.6189&	0.5978&	0.5868   \\
          &       &       & DICE  & 0.6340  & 0.6137  & 0.5769  & 0.5616  & 0.5236  & 0.5057  \\
          &       &       & EDA*  & \textbf{0.5885} & \textbf{0.5846} & \textbf{0.5663} & \textbf{0.5375} & \textbf{0.5049} & \textbf{0.5016} \\
\cline{2-10}          & \multirow{6}[3]{*}{EM} & \multirow{6}[3]{*}{NMI} & Unattack & 0.4968  & 0.4968  & 0.4968  & 0.4968  & 0.4968  & 0.4968  \\
          &       &       & RA    & 0.4961  & 0.4926  & 0.4913  & 0.4860  & 0.4867  & 0.4851  \\
          &       &       & RLS    & 0.4970&	0.4950&	0.4916&	0.4915&	0.4917&	0.4830   \\
          &       &       & DBA    & 0.4894&	0.4847&	0.4872&	0.4895&	0.4896&	0.4932   \\
          &       &       & DICE  & 0.4829  & 0.4780  & 0.4608  & 0.4525  & 0.4467  & 0.4368  \\
          &       &       & EDA*  & \textbf{0.4478} & \textbf{0.4518} & \textbf{0.4462} & \textbf{0.4336} & \textbf{0.4187} & \textbf{0.4091} \\
          \hline
    \bottomrule
    \end{tabular}%
  \label{tab:addlabel}%
\end{table*}%

\begin{figure}[!t]
        \centering
            \subfigure[Added links]{
                \includegraphics[width=1.1\linewidth]{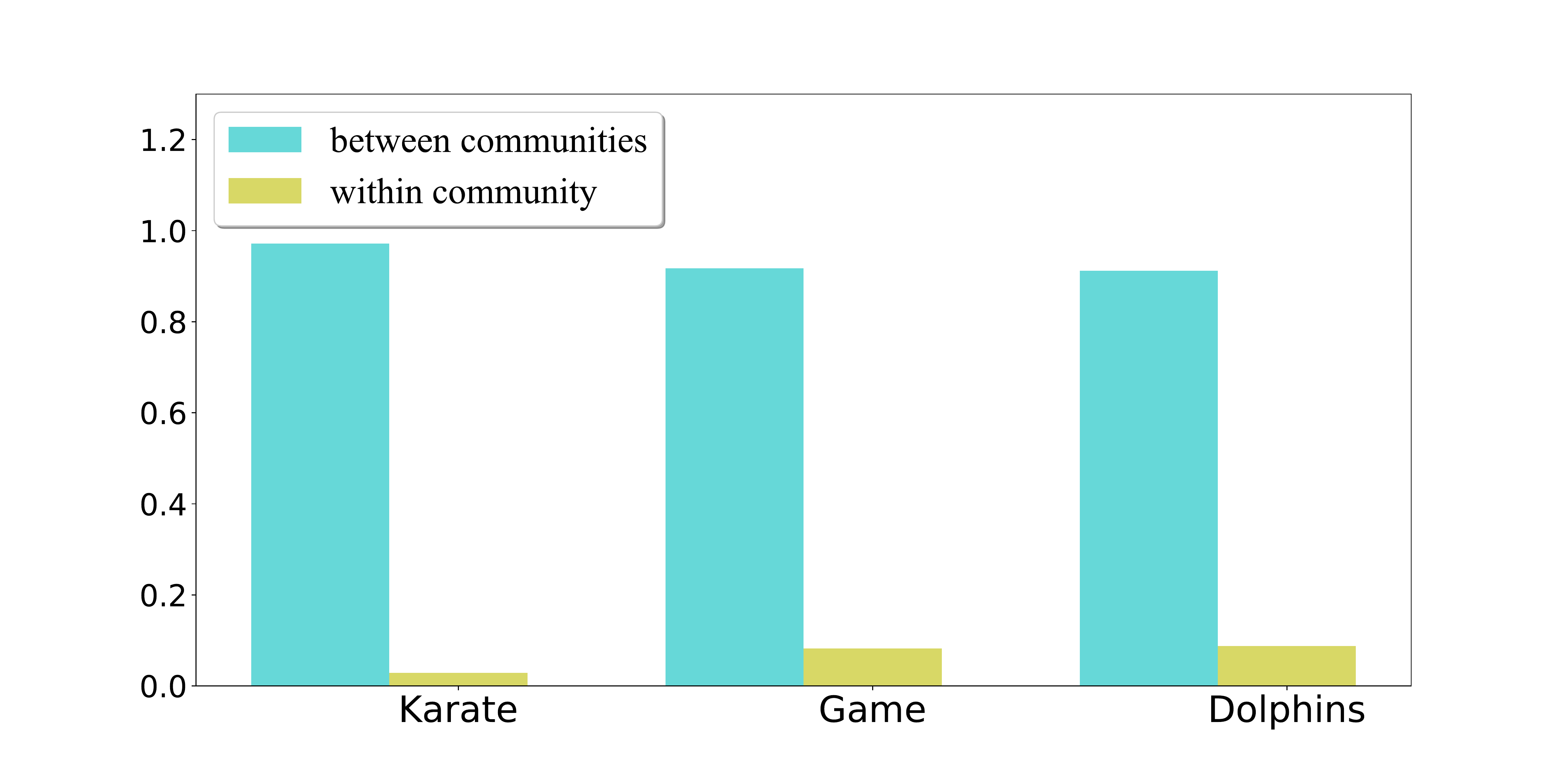}
            }
            \subfigure[Deleted links]{
                \includegraphics[width=1.1\linewidth]{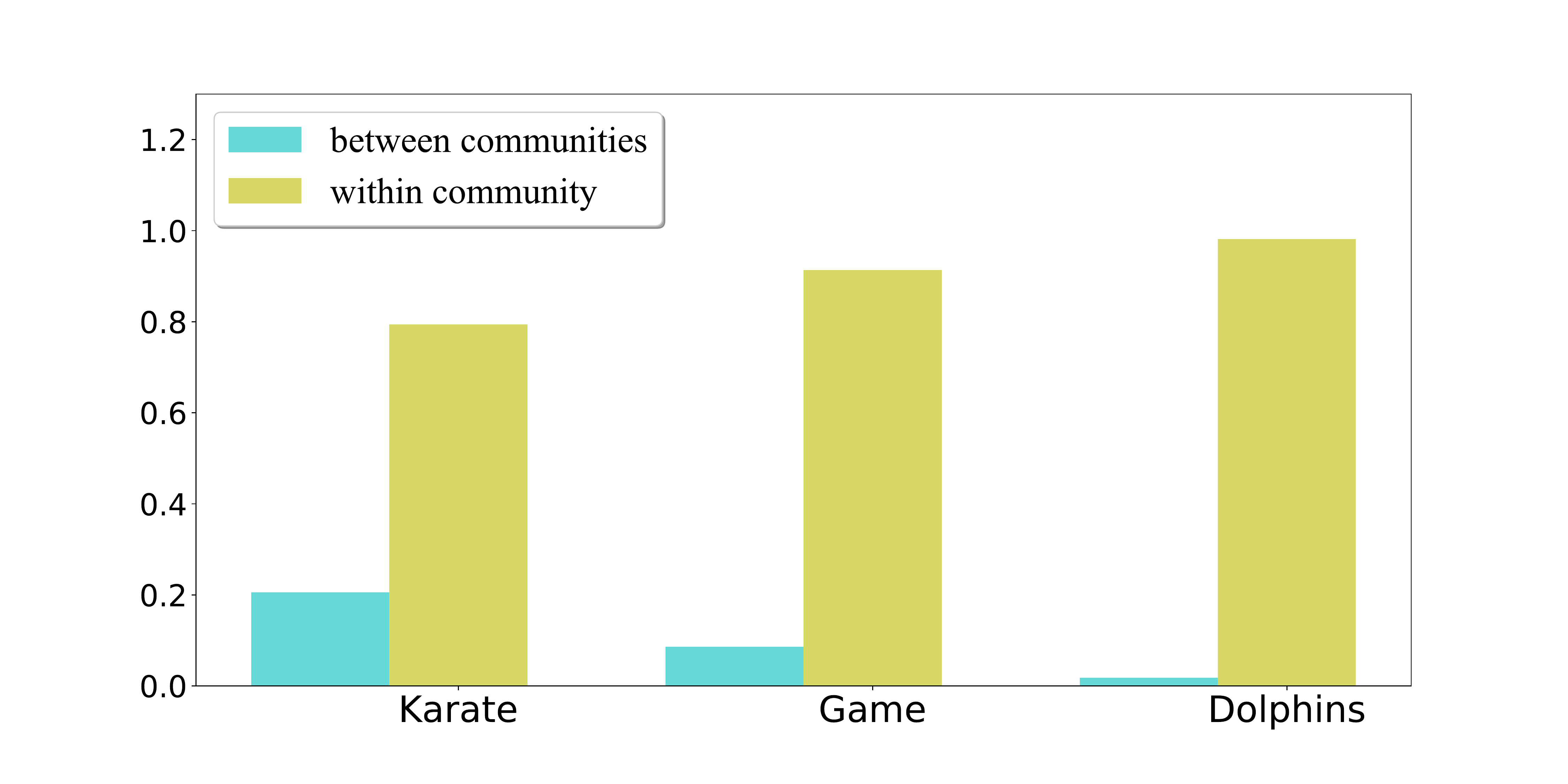}
            }
        \caption{The percentages of (a) added links and (b) deleted links within community and between communities.}
        \label{Fig:network}
\end{figure}

\subsection{Transferability of EDA}
Generally, disturbing the distance matrix between node vectors in embedding space is equivalent to altering the similarity between nodes in the network, which will naturally affect other algorithms. We thus also examine the transferability of the proposed DeepWalk-based EDA method for other network algorithms.

In particular, we choose another network embedding method High-Order Proximity preserved Embedding (HOPE)~\cite{ou2016asymmetric} and two classic network algorithms, including Label Propagation Algorithm (LPA)~\cite{raghavan2007near} and Eigenvectors of Matrices (EM)~\cite{newman2006modularity}, which are not based on network embedding. HOPE utilizes the generalized SVD to handle the formulation of a class of high-order proximity measurements. HOPE is widely used due to its high effectiveness and efficiency. LPA sets the label of a node identical to most of its neighbors through an iterative process, while EM directly uses the modularity matrix to study the community structure. We choose the two relatively large networks, i.e., Game and cora, to study the transferability of EDA to avoid a cross-page table TABLE~\ref{tab:addlabel}.

The results are shown in TABLE~\ref{tab:addlabel}. Although EDA is based on DeepWalk, it is still valid on HOPE-based node classification and community detection algorithms, i.e., HOPE+LR and HOPE+K-Means. Moreover, it is also effective on LPA and EM, which are not based on any network embedding algorithm. EDA still outperforms the other baseline attack strategies in most cases, suggesting that it has relatively strong transferability, i.e., we can generate small perturbations on the target network by EDA and destroy many network algorithms, no matter whether it is based on a certain network embedding method.


 \subsection{Visualization and explanation}
\textbf{Statistics of flipping links.}
To better understand how EDA works in reality, we visualize the perturbations of the added and deleted links to see how many of them are within the same communities or between different communities. Taking the karate club network as an example, we present its original network and the adversarial network after one-link attack, as shown in Fig.~\ref{Fig:net-attack}, where we can see that the added link is between two different communities while the deleted one is within the same community. To give more comprehensive results, we consider all the flipping links in all the experiments for experimental networks, and count the percentages of added links and deleted links within or between communities, respectively, as shown in Fig.~\ref{Fig:network}. We find that most of the added links are between communities, while most of the deleted links are within the same community. This rule is quite interesting since EDA focuses on perturbing the network from embedding space without any prior knowledge of communities. One reason may be that the community is a critical structural property that matters the embedding results, and this is why EDA behaves quite well on attacking community detection algorithms.

\textbf{The position of node vector.}
Furthermore, to show what EDA actually does to network embedding, we also visualize node embedding vectors by t-sne method after the attack.
For different percentages of rewiring links ranging from 1\% to 7\%  in karate club network, we obtain the nodes vectors of original and adversarial networks using DeepWalk, and then display the results in a two-dimensional space, as shown in Fig.~\ref{Fig:clustervis}. There are four communities represented by different colors. When there is no attack, the node vectors of different communities are separated, as shown in Fig.~\ref{Fig:clustervis} (a). As the number of rewiring links increases, the node vectors of different communities are being mixed gradually, and finally become inseparable, as shown in Fig.~\ref{Fig:clustervis} (h). This result demonstrates that EDA indeed has a significant effect on network embedding, and can further disturb the subsequent community detection or node classification effectively.

 \section{Conclusion}
 \label{sec:conclusion}
In this paper, we propose the novel unsupervised attack strategy, namely EDA, on network embedding, with the focus on disturbing the Euclidean distance between nodes in the embedding space as much as possible with minimal changes of the network structure. Since network embedding is becoming the basis of many network algorithms, EDA can be considered as a universal attack strategy that can degenerate many downstream network algorithms.

We take the DeepWalk as the basis to design our EDA, and a number of experiments on real networks validate that EDA is more effective than a set of baseline attack strategies on community detection and node classification, no matter whether these algorithms are based on DeepWalk or other embedding methods, or even not based on embedding. Such results indicate the strong transferability of our strategy. Not that, our EDA is an attack on disturbing global network structure, and we may also focus on disturbing local structure around target nodes and links, to realize target attacks, which belongs to our future work.

\bibliographystyle{IEEEtran}
\bibliography{EDA}

\end{document}